\documentclass[a4paper,fleqn]{cas-sc}
\usepackage[numbers]{natbib}
\usepackage{subfigure}
\usepackage[ruled]{algorithm2e}
\usepackage{bm}

\usepackage{caption}
\usepackage{amssymb}
\usepackage{graphicx}
\usepackage{float} 

\usepackage{makecell}
\usepackage{multirow}

\usepackage{cleveref}
\usepackage{color}

\usepackage{hyperref}
\hypersetup{hypertex=true,
            colorlinks=true,
            linkcolor=blue,
            anchorcolor=blue,
            citecolor=blue}

\def\tsc#1{\csdef{#1}{\textsc{\lowercase{#1}}\xspace}}
\tsc{WGM}
\tsc{QE}
\tsc{EP}
\tsc{PMS}
\tsc{BEC}
\tsc{DE}

\usepackage{caption}
\begin{document}
\let\WriteBookmarks\relax
\def\floatpagepagefraction{1}
\def\textpagefraction{.001}
\let\printorcid\relax
\shortauthors{Linlin Wang et~al.}

\title [mode = title]{Traffic Flow and Speed Monitoring Based On Optical Fiber Distributed Acoustic Sensor}                  
\tnotemark[1]

\author[1]{Linlin Wang}[style=chinese]
\credit{Conceptualization, Methodology, Formal analysis, Visualization, Writing - original draft}
\address[1]{School of Mathematics, Renmin University of China, Beijing 100872, China}

\author[1]{Shixin Wang}[style=chinese]
\credit{Investigation, Software, Validation}

\author[2]{Peng Wang}[style=chinese]
\credit{Writing - review \& editing}
\address[2]{School of Information, Renmin University of China, Beijing 100872, China}

\author[1]{Wei Wang}[style=chinese]
\credit{Methodology, Validation, Writing - review \& editing}

\author[3]{Dezhao Wang}[style=chinese]
\credit{Conceptualization, Data curation}
\address[3]{Beijing Jhbf Technology Development Co., Ltd., China}

\author[2]{Yongcai Wang}[style=chinese]
\credit{Writing - review \& editing, Validation, Supervision}

\author[1]{Shanwen Wang}[style=chinese]
\cormark[1]
\credit{Conceptualization, Project administration, Funding acquisition, Supervision, Writing - review \& editing}
\ead{s_wang@ruc.edu.cn}

\tnotetext[1]{Linlin Wang, Shixin Wang, Wei Wang, and Shanwen Wang were funded by Tianjin Yunhong Technology Development Grant No. 2021020531. Peng Wang and Yongcai Wang received support from the National Natural Science Foundation of China Grant No. 61972404.}

\cortext[1]{Corresponding author.}

\begin{abstract}
    In the realm of intelligent transportation systems, accurate and reliable traffic monitoring is crucial. Traditional devices, such as cameras and lidars, face limitations in adverse weather conditions and complex traffic scenarios, prompting the need for more resilient technologies. 
    This paper presents traffic flow monitoring method using optical fiber-based distributed acoustic sensors (DAS). An innovative vehicle trajectory extraction algorithm is proposed to derive traffic flow statistics. In the processing of optical fiber waterfall diagrams, Butterworth low-pass filter and peaks location search algorithm are employed to determine the entry position of vehicles. Subsequently, line-by-line matching algorithm is proposed to effectively track the trajectories. Experiments were conducted in highway, tunnel and city scenarios. Visualizations show that our approach not only extracts vehicle trajectories more accurately than the classical Hough and Radon transform-based methods and MUSIC beamforming algorithm, but also facilitates the calculation of traffic flow information using the low-cost acoustic sensors. It provides a new reliable means for traffic flow monitoring which can be integrated with existing methods like vision-based method. 
\end{abstract}


\maketitle

\section{Introduction}

With the advancing opportunities in technology, intelligent transportation systems (ITS) have emerged as a promising direction for future transportation
systems \cite{figueiredo2001towards}. These systems can aid in monitoring and managing traffic flow using various tools such as cameras and sensors. Such technologies facilitate real-time traffic monitoring and control, reducing congestion and enhancing traffic flow \cite{2023intelligent}. Central to these are the components of vehicle detection and trajectory extraction. Vehicle detection involves recognizing and localizing target objects \cite{azimjonov2021real}. Vehicle trajectories, on the other hand, are derived from vehicle detection and tracking data gathered by devices on the road. Given the motion trajectory and environmental changes, accurately predicting a vehicle's future trends and behaviors becomes vital for traffic monitoring \cite{wang2023multimodal}. Nonetheless, the efficient detection of features and precise prediction of a target vehicle's trajectory remains challenging.

\begin{figure}[!t]
    \centering
    \subfigure[]{\includegraphics[width=4in]{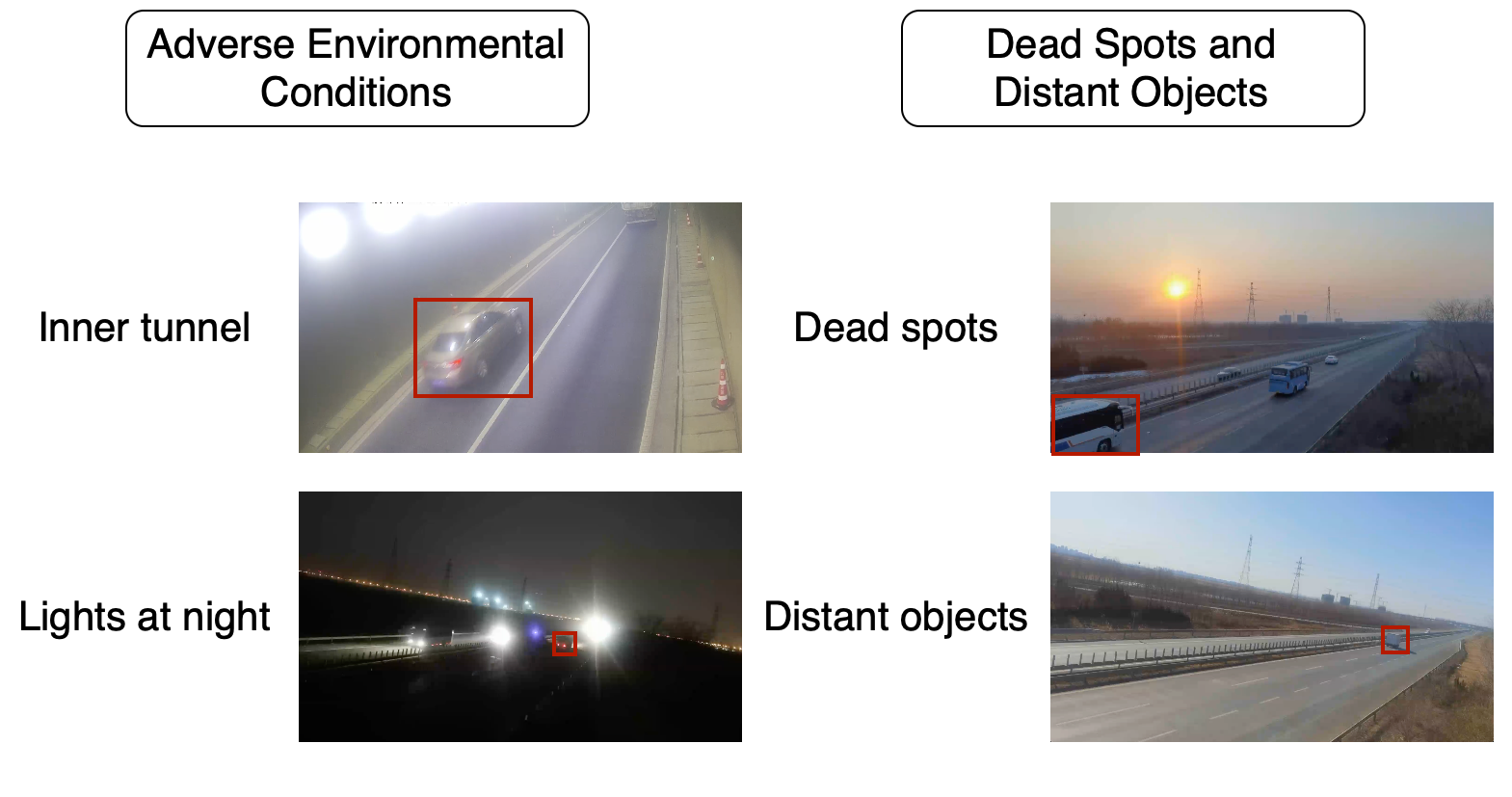}}
    \hfill
    \subfigure[]{\includegraphics[width=4in]{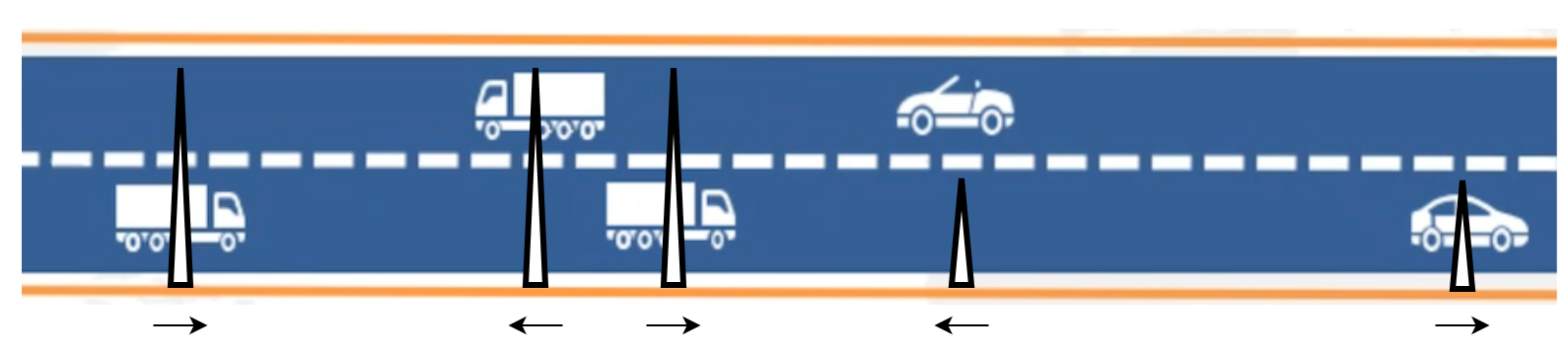}}
    \hfill
    \caption{Comparison of vision-based and fiber-based systems. (a) Traffic monitoring under environmental conditions that cameras constraints. (b) A schematic diagram of a multi-lane highway using optical fiber systems.}
    \label{fiber}
\end{figure}

Nowadays, methods for monitoring traffic flow have predominantly turned to technologies such as cameras, radars, and lidars \cite{guerrero2018sensor}. Deep Learning algorithms, including YOLO (2D multi-object detection) \cite{wang2022yolov7} and CenterPoint (3D multi-object detection) \cite{yin2021center}, are increasingly utilized for vehicle detection. Additionally, an effective combination of the Kalman filter and the Hungarian algorithm can be harnessed for state estimation and data association, thereby facilitating vehicle tracking and real-time positioning \cite{bewley2016simple}. Nonetheless, as shown in Fig. \ref{fiber}(a), the efficacy of these devices can be severely compromised by environmental conditions and deployment constraints. Compounding these issues, cameras, given their fixed viewing angles, introduce dead zones and issues relating to object sizes \cite{moon2002performance}. These problems are accentuated when detecting smaller objects nearby or larger objects at extended distances, as illustrated in Fig \ref{fiber} (a). Present 2D detection methods often fall short in identifying distant small objects. While lidars can discern distance information over longer ranges, their high costs hinder widespread deployment, especially at highway ends \cite{williams2013synthesis}. Contrasting with cameras and lidars, which serve as “visual detectors”, underground optical fibers can be likened to “acoustic detectors”. As shown in Fig \ref{fiber} (b), when a vehicle passes by, the optical fiber records its location coordinates in the form of a wavelet, which is not affected by adverse environmental conditions, and there are no dead spots. Benefiting from numerous design advantages such as resistance to electromagnetic interference, heightened sensitivity, cost-effectiveness, durability, stealth, and adaptability, optical fibers may provide another reliable way for continuous and comprehensive traffic state perception in smart highways \cite{wang2019comprehensive} and can be integrated with camera and Lidar-based methods.

In this study, we investigate optical fiber-based distributed acoustic sensing (DAS) for monitoring traffic in highway, tunnel and city scenarios. The proposed methods mainly include two parts: (1) vehicle detection and tracking by signal processing; and (2) traffic density, flow, and velocity estimation. After preprocessing optical fiber waterfall diagrams, we propose a vehicle position detection algorithm based on the Butterworth low-pass filter and peaks location search. Then vehicle trajectories are extracted based on line-by-line matching and polynomial fitting, thus achieving real-time detection and overall tracking for vehicles. We then present the calculation methods for three traffic flow information (i.e. velocity, flow and density) on road profile and road segment respectively, realizing traffic flow and speed monitoring.

The performances of our proposed algorithms are verified in real datasets. We use self-established dataset for highway and tunnel scenarios, and public dataset from \cite{van2022deep} for city scenario. The vehicle tracking results extracted by our methods are more accurate than the traditional transform-based Hough \cite{wiesmeyr2021distributed, catalano2021automatic}, Radon \cite{wang2021vehicle} and MUSIC beamforming method \cite{van2022deep}. In the tunnel scenario, our algorithm accuracy is up to 90.90\%, which validates the usability of the proposed methods in practical scenarios considering the very low-cost signal processing-based method. Even facing complex vehicle behaviors like congestion and illegal overtaking, our proposed methods can track vehicles when lines overlap or intersect, validating the effectiveness of the proposed methods for complicated vehicle behaviors.

In this article, we explore the applicability of DAS systems in traffic monitoring and the main contributions are as follows:

\begin{enumerate}

    \item To the best of our knowledge, we have pioneered the application of DAS systems in real tunnel scenario.
    
    \item A novel position detection algorithm and a trajectory extortion algorithm are proposed for vehicle detection and tracking, relying solely on optical fiber signals. 

    \item Experiments in highway, tunnel and city scenarios demonstrate the effectiveness of the proposed approach, even when facing challenges like vehicle overtaking and congestion-induced crossover situations.

\end{enumerate}

\section{Related Works}
\label{Related_Work}
Since DAS systems have been utilized in transportation \cite{huang2019first}, numerous exploratory efforts are emerging for traffic monitoring. In this section, we introduce theoretical analyses, existing algorithms, and application scenarios of DAS systems.
\subsection{Theoretical Analysis}

In our experiments, fibers were buried underground at a certain depth on one side of the road. When a vehicle passes by, the vehicle's pressure deforms the road surface and is transmitted to the optical fiber sensor, generating a strain that can be measured for amplitude. This geodetic deformation can be depicted by the Flamant-Boussinesq model \cite{jousset2018dynamic}.

Denote a given position in three-dimensional space as $(d_x, d_y, d_z)$, where $d_x$ and $d_y$ are the tangential and perpendicular distances to the road respectively, and $d_z$ is laying depth. The quasi-static or geodetic deformation $P(d_x,d_y,d_z)$ at this position is:
\begin{equation}
    P(d_x,d_y,d_z) = \frac{F}{4 \pi G} \times p(d_x,d_y,d_z),
\end{equation}
where $F$ is the total force onto an infinite half-space, $G$ is the uniform shear modulus, and $p(d_x,d_y,d_z) = \frac{d_x}{r^2}(\frac{d_z}{r} + \frac{2 \nu - 1}{1 + d_z/r})$ with $r = \sqrt{d_x^2+d_y^2+d_z^2}$ and Poisson's ratio $\nu$. 

Consequently, the impulse response $k$ of DAS systems in response to a vehicle passing by a given DAS sensor can be expressed as:
\begin{equation}
    k = \frac{1}{l}\left[P(d_x+\frac{l}{2}, d_y, d_z) - P(d_x-\frac{l}{2}, d_y, d_z)\right],
\end{equation}
where $l$ is the DAS gauge length. 

\begin{figure}[t!]
    \centering
    \subfigure[$d_y$ change]{
        \includegraphics[width=2.5in]{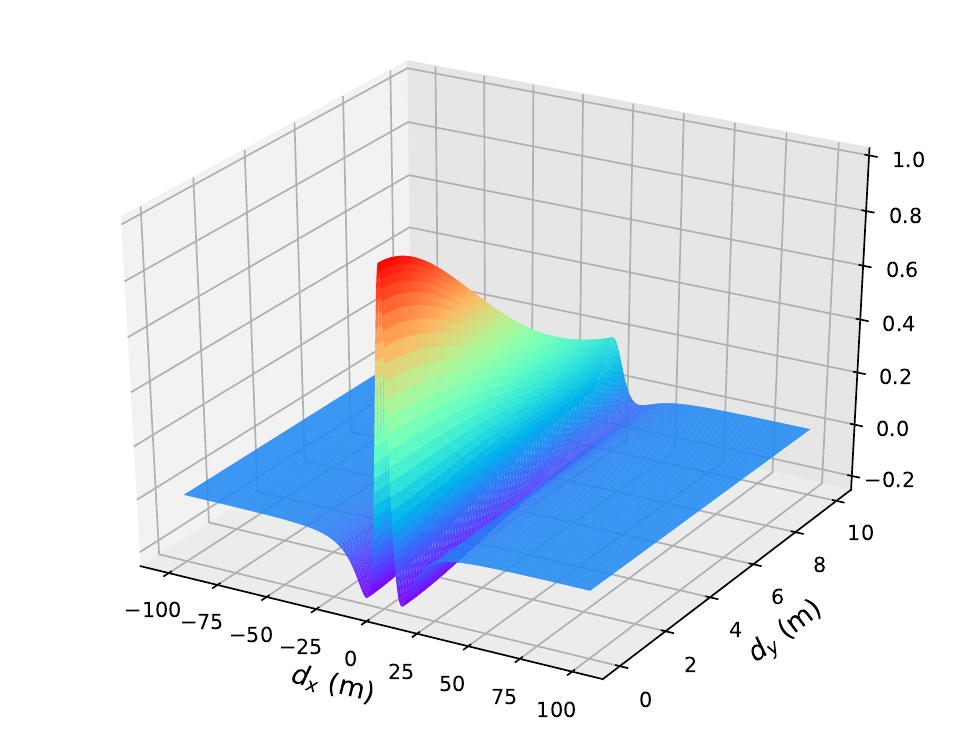}
    }
    \subfigure[$F$ change]{
        \includegraphics[width=2.5in]{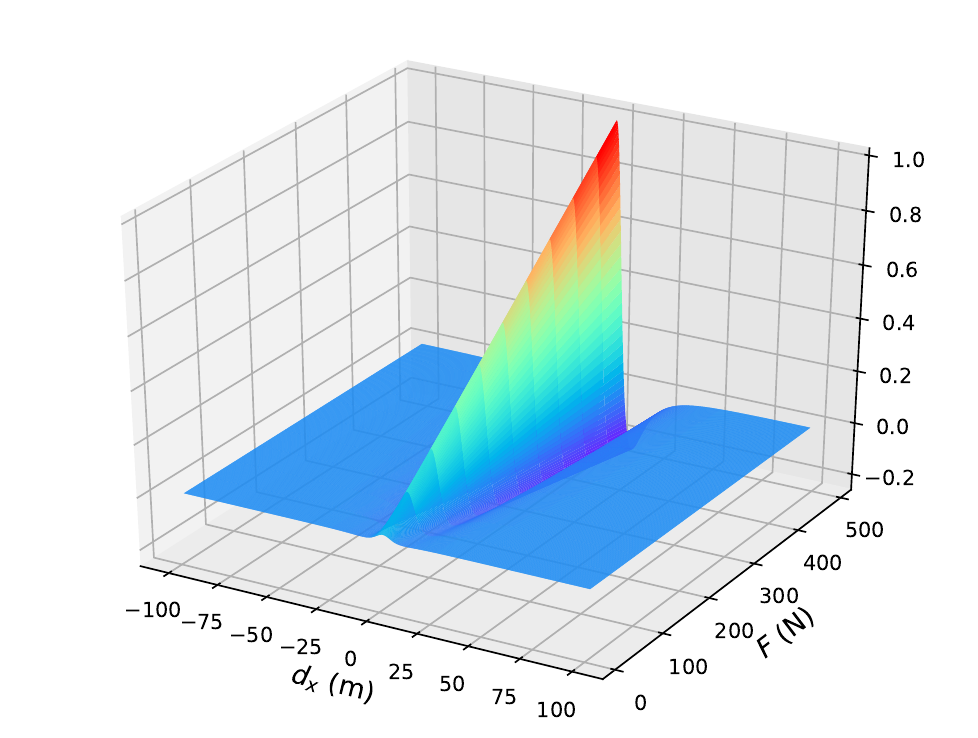}
    }
\caption{Numerical simulation of the quasi-static signals. (a) Various road-perpendicular offsets, $d_y$, to the fiber, (b) various vehicle loads $F$.}
\label{F-B}
\end{figure}

It can be seen from Fig. \ref{F-B} that theoretical amplitudes of signals decrease with the road-perpendicular offsets, $d_y$ and increase with the vehicle loads, $F$. 

In the actual calculation process, each vehicle has four wheel positions that generate geodetic deformation. Assuming the vehicle's front track and length as $a$ and $b$ respectively, we can then calculate the geodetic deformation at each of the four wheels. Impulse response $k$ can be defined as the difference between the response $k_{x_1}$ at the front end of the vehicle $d_{x_1} = d_x + \frac{l}{2}$ and the response $k_{x_2}$ at the rear end of the vehicle $d_{x_2} = d_x - \frac{l}{2}$:
\begin{equation}
   k = |k_{x_2} - k_{x_1}|,
\label{impulse}
\end{equation}
where
\begin{equation}
    k_{x_1} = \sum_{i=1}^{4} w_i \cdot P(d_{x_1} + \alpha_i, d_y + \beta_i, d_z),
    \end{equation}
\begin{equation}
    k_{x_2} = \sum_{i=1}^{4} w_i \cdot P(d_{x_2} + \alpha_i, d_y + \beta_i, d_z).
\end{equation}

Here, $w_i$ represents the weight of the $i$-th wheel, and $(\alpha_i, \beta_i)$ represents the coordinates of the $i$-th wheel relative to the vehicle center. From left front to right rear of a vehicle, $(\alpha_i, \beta_i)$ are $(\frac{b}{2}, \frac{a}{2}), (\frac{b}{2}, -\frac{a}{2}), (-\frac{b}{2}, -\frac{a}{2}) (-\frac{b}{2}, \frac{a}{2})$.

\begin{figure}[t!]
    \centering
    \subfigure[Different model with different $d_y$.]{
        \includegraphics[width=2.5in]{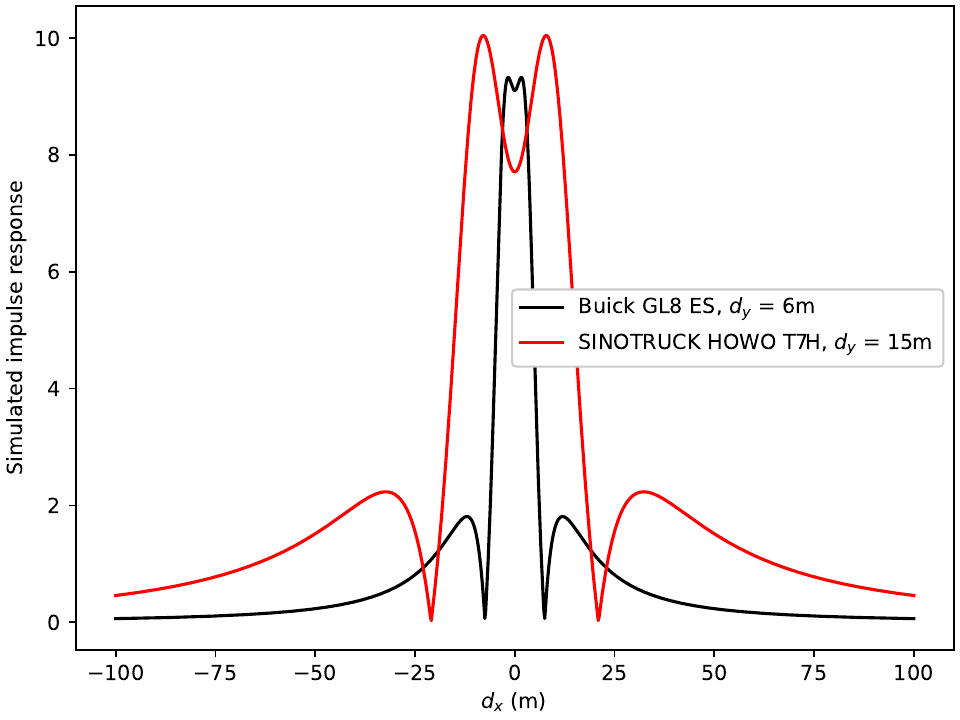}
    }
    \subfigure[Different length with same weight.]{
        \includegraphics[width=2.5in]{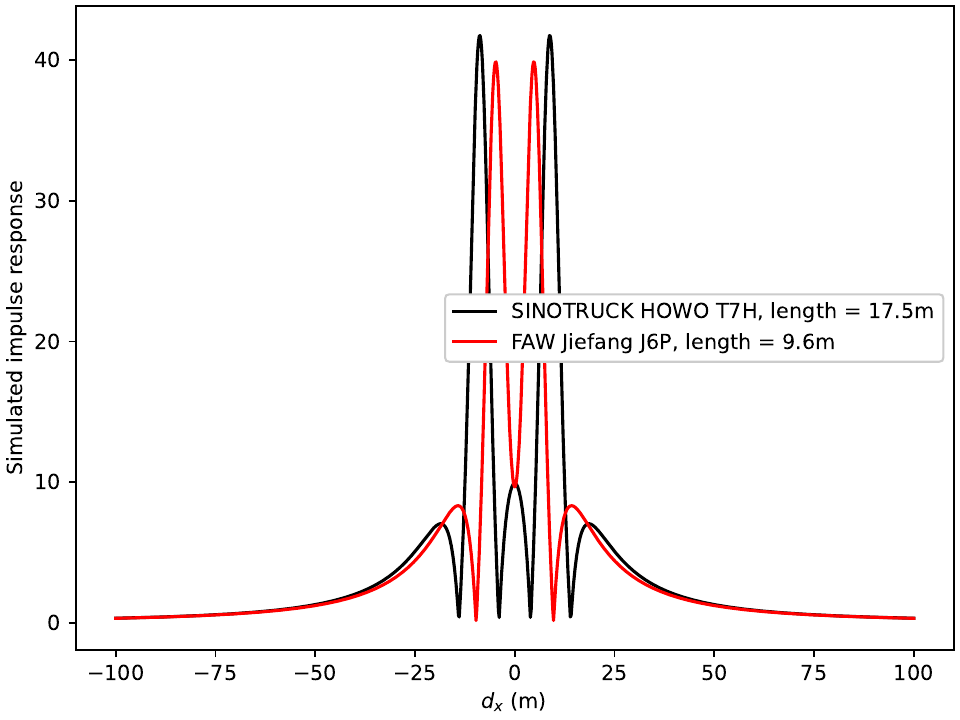}
    }
\caption{Two instances that can result in similar waveforms. It is worth noting that vehicles with longer lengths show double peaks, which is also consistent with the real situation in Fig. \ref{track} (a) in our data set.}
\label{similar_waveforms}
\end{figure}

Some similar complex phenomena are also discussed in this section. We compared the simulated wave effect of a popular luxury MPV Buick GL8 ES (5.3m in length, 2.5t in load) with that of SINOTRUCK HOWO T7H (17.5m in length, 20t in load) designed for long-distance transportation, as shown in Fig. \ref{similar_waveforms}(a). Furthermore, we chose SINOTRUCK HOWO T7H (17.5m in length) and FAW Jiefang J6P (9.6 in length) of same weight for comparison. These two cases show similar amplitude information, which is also the problems for fiber signal acquisition to identify vehicle models.

During actual measurements, challenges arise such as variations in optical fiber burial depths, changes in soil shear modulus, and mutual interference of vehicle signals on the road, introducing more complexities.

\subsection{Existing denoising and preprocessing algorithms}

Due to constraints in data acquisition, transmission, storage and other factors, fiber signals often contain significant amounts of noise, necessitating advanced denoising processing. Due to the advantages of wavelet theory in both time and frequency domains, a series of denoising algorithms are designed, such as a six-level wavelet multi-scale decomposition algorithm \cite{wu2015separation}, continuous wavelet transform \cite{huot2017automatic}, a combination of soft and hard threshold wavelet denoising algorithm \cite{liu2018traffic, an2023traffic} and wavelet threshold denoising \cite{hou2024method}. However, these complex improved algorithms based on wavelets slow down the response speed.

In recent years, machine learning and deep learning have demonstrated significant accomplishments. Filtering algorithms and SVM are utilized for signal denoising and classification \cite{liu2019vehicle, min2024vehicle}. An attention-based convolutional neural network architecture has been specifically devised for denoising DAS signals \cite{wang2021rapid}. Time-DAE \cite{van2022deep} and Spatial-DAE \cite{yuan2023spatial}, based on deconvolution auto-encoder, have been proposed to deconvolve signals with impulse responses. Several kinds of novel DL methods including AlexNet, SVGG and SR-Net are designed to process DAS signals \cite{zhong2024distributed}. Though these algorithms can facilitate subsequent processing workflows, deep learning algorithms lack mobility and interpretability, and there are few public data sets available.

\subsection{Existing vehicle detection and tracking algorithms}

In the process of vehicle feature recognition and trajectory extraction, researchers often treat vehicle trajectories as straight lines and conduct domain transformations to identify the extreme points in the transformation domain. Various approaches like Hough transform \cite{wiesmeyr2021distributed, catalano2021automatic} and its improvement \cite{corera2023long}, Radon transform \cite{wang2021vehicle}, and the MUSIC beamforming algorithm \cite{van2022deep} are commonly employed for this purpose. However, these methods are primarily suited for extracting lines and may not be efficient for monitoring changes in vehicle positions and real-time tracking.

To address this limitation, an improved double threshold algorithm is utilized for vehicle detection \cite{liu2019vehicle}. Additionally, a K-means clustering algorithm and Kalman filtering technology \cite{thulasiramantraffic} are employed for vehicle identification and tracking, with algorithm verification conducted on a simple traffic flow data subset. Furthermore, a spatial Bayesian filtering and smoothing algorithm are developed for detecting, tracking, and characterizing individual vehicles \cite{liu2023telecomtm}. However, these algorithms tend to perform well on roads with relatively low traffic volume and uncomplicated traffic patterns.

\subsection{Application scenarios in transportation}

In the application scenario of DAS systems in traffic monitoring, researchers often begin with simple scenarios where vehicles pass successively, presenting relatively straightforward situations in various settings such as smart cities \cite{hall2019using}, Nanshan Iron Mine \cite{liu2018traffic}, villages \cite{van2022deep}, Brady Hot Springs \cite{chambers2020using}, San Jose, California \cite{liu2023telecomtm}, Pasadena, California \cite{wang2021ground} during COVID-19 pandemic, as well as on campuses like Stanford campus \cite{huot2017automatic} and Beijing Jiaotong University \cite{liu2019vehicle}. Some researchers have also explored highway scenarios, including studies like those in \cite{wang2021rapid} and \cite{wiesmeyr2021distributed}. However, these configurations overlook challenges posed by long-distance installations and continuous tracking of vehicles. To the best of our knowledge, no studies have been conducted on DAS systems for traffic monitoring in tunnels. Besides, there have been no studies on complex vehicle behaviors such as congestion and overtaking. These scenarios present challenges within the realm of optical fiber-based traffic monitoring systems.

\section{Method}
\label{Method}
\begin{figure}[!t]
    \centering
    \includegraphics[width=6in]{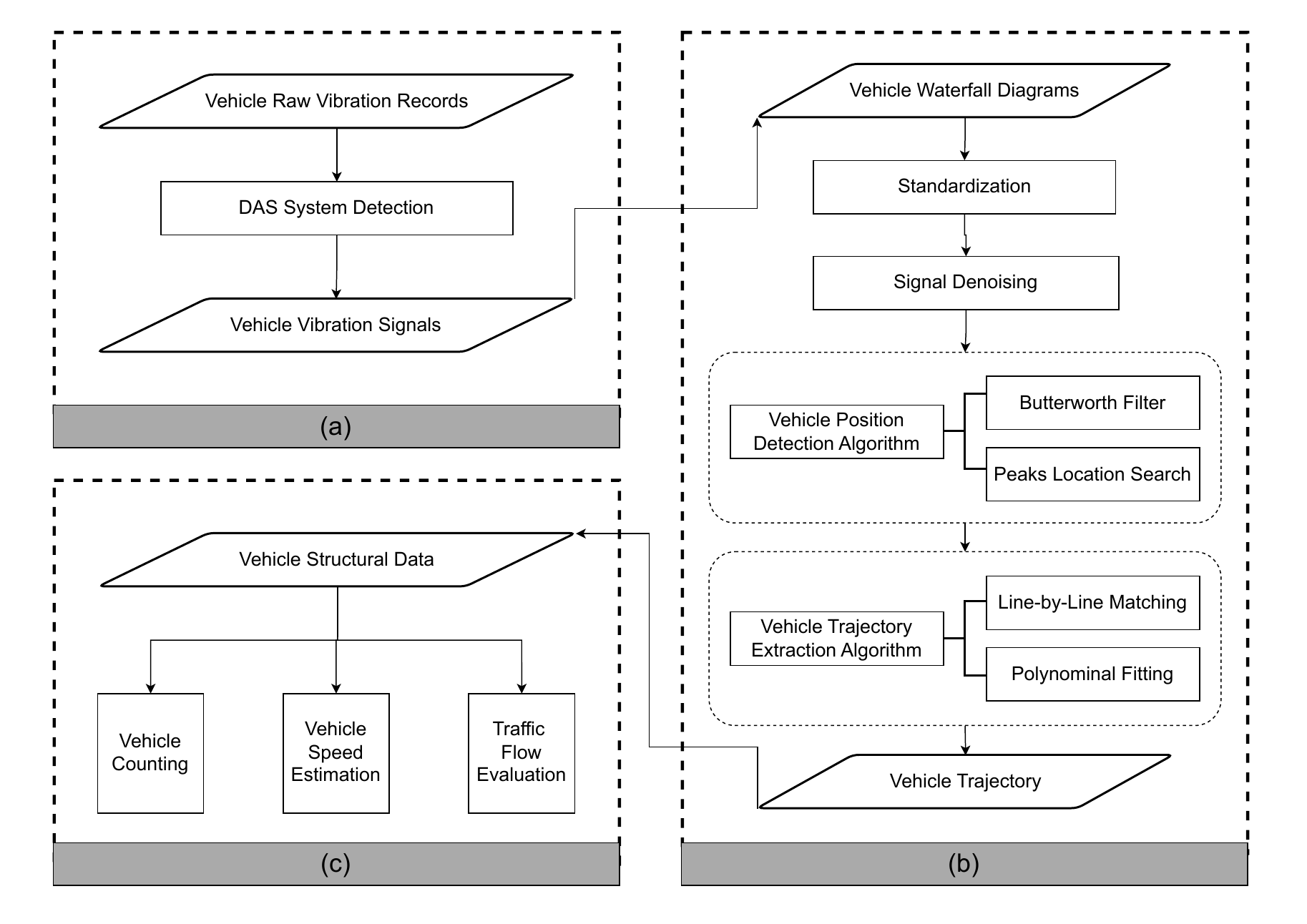}
    \caption{Workflow of the total algorithm. (a) Vehicle vibration data detected by DAS systems, (b) signal processing for vehicle localization and trajectory extraction, and (c) structural data for traffic monitoring.}
    \label{workflow}
\end{figure}

In this section, we formulate the problem and propose a vehicle position detection algorithm and vehicle trajectory extraction algorithm based on optical fiber signals after preprocessing. The methodology is outlined in Fig. \ref{workflow}.

\subsection{Problem Formulation}

The original optical fiber data recorded by DAS systems can be represented as a matrix denoted by A:
\begin{equation}
A = \begin{pmatrix}
	 a_{11} & a_{12} & \cdots & a_{1n} \\
	a_{21} & a_{22} & \cdots & a_{2n} \\
    \vdots & \vdots & \ddots & \vdots \\
	a_{m1} & a_{m2} & \cdots & a_{mn}
	 \end{pmatrix} = (a_{ij})
\label{matrix}
\end{equation}
where $m$ denotes the time points, $n$ denotes the distance points, and $a_{ij} \in \mathbb{R^{+}}$. To facilitate the visualization of vehicle data collected from optical fiber, we can draw optical fiber waterfall diagrams and display fiber stripes:

\textbf{Fiber waterfall diagram:} Waterfall diagrams after visualization of fiber matrix A in Eq. (\ref{matrix}) with distance as the horizontal axis and time as the vertical axis.

\textbf{Fiber stripe:} Clear banded stripes visible to human eyes in a fiber waterfall diagram. In this paper, fiber stripes represent the driving trajectories of the vehicles.

The primary objective of this study is to validate the effectiveness of DAS systems as a novel traffic monitoring method. Additionally, accurately distinguishing vehicle trajectories and conducting traffic flow statistics under complex conditions such as congestion and overtaking poses a challenge that needs to be addressed.

\subsection{Signal Preprocessing}

In signal processing, denoising is the primary task. After the min-max normalization, a wavelet threshold denoising algorithm combining hard and soft threshold functions is employed. 

Due to its excellent time-frequency characteristics, wavelet theory has gained widespread use in signal processing and has experienced rapid development \cite{antonini1992image}. Signal energy correlates with high-amplitude wavelet coefficients, and noise energy correlates with low-amplitude wavelet coefficients. An optimal threshold can be set to identify valuable signals above it for retention or contraction, while coefficients below it are considered noise and eliminated. Then signals are reconstructed using inverse wavelet transform to obtain a denoised signal. For the selection of the wavelet threshold function, a wavelet threshold denoising algorithm combining hard and soft threshold functions is employed to denoise vehicle vibration signals. The formula is as follows:
\begin{equation}
    \overline{{\omega}}_{j,k}=\begin{cases}
    {\omega}_{j,k} - a \lambda,& {\omega}_{j,k} \geq \lambda;\\
    0,& \lvert {\omega}_{j,k} \rvert < \lambda;\\
    {\omega}_{j,k} + a \lambda,& {\omega}_{j,k} \leq -\lambda,\end{cases}
\label{wavelet}
\end{equation}
where ${\omega}_{j,k}$ is the wavelet coefficient and $\lambda$ is the threshold. In this paper, according to the actual situation, an ideal denoised result can be obtained when $a = 0.5$.

\subsection{Vehicle Position Detection}
We propose an algorithm for vehicle detection based on the Butterworth filter and peaks location search algorithm. The vehicle position detection algorithm flow is shown in Algorithm \ref{algorithm_1}.  

\begin{algorithm}[htbp]
    \caption{Position Detection Algorithm.}
    \label{algorithm_1}
    \LinesNumbered 
    \KwIn{$A = (a_{ij})$: original optical fiber matrix;\ $i = 1, ..., m$, $j = 1, ..., n$, $a_{ij} \in \mathbb{R^{+}}$: selected time and distance.}
  
    \KwOut{$D = (d_{ij})$: preproccessed optical fiber matrix; $S$: Vehicle counts that can be extracted; $k_s \in \mathbb{N^{+}}, s = 1, ...,S$: appearance time of each vehicle.}
    $B = (b_{ij})$ $\gets$ MinMaxNormalization($A =(a_{ij})$)\;
    $D = (d_{ij})$ $\gets$ WaveletDenoising($B = (b_{ij})$)\;
    $k_s$ $\gets$ FindPeak(ButterworthFilter($D_{1: m,1}$))\;
    $(k_s, 1)$ is the first pair of time and distance position points of each extracted vehicle trajectory\;
  
  \textbf{Return:} Vehicle counts $S$; appearance time of each vehicle $k_s$; preprocessed optical fiber matrix $D = (d_{ij})$.
\end{algorithm}

\subsubsection{Smoothing method of Butterworth Filter}

For optical fiber waterfall diagrams after preprocessing, we further adopt Butterworth low-pass filtering on the signals at the starting position of optical fibers. Butterworth filter is a low-pass filter with maximum flat amplitude response in the pass-band \cite{butterworth1930theory}. Compared with analog filter, Butterworth filter has many advantages such as high precision, stability, flexibility, and no impedance matching \cite{winder2002analog}. Butterworth Filter has been widely used in many fields such as automatic control, image and communication. The Butterworth filter has two important parameters, order $N$ and normalized cut-off frequency $W_n$. The order determines the steepness of the frequency response curve, and the cut-off frequency is usually used to control the frequency characteristics. In optical fiber signal processing, noise mostly presents high frequency characteristics, so we choose Butterworth low-pass filter to remove the high frequency noise in the image and smooth signal details. In this way, we can further separate components of different frequencies, so as to make the signals even flatter.

\subsubsection{Peaks location search algorithm}

The features of wave peaks satisfy the principle that the first derivative is zero and the second derivative is negative, the position of the wave peaks can be searched as follows:

(1) Define the projection curve as $V = [v_1, v_2, \ldots, v_n]$. 

(2) Calculate the first-order difference vector ${\rm {Diff}}_v(i)$ for $i \in 1, 2, \ldots , N-1$:
\begin{equation}
{\rm {Diff}}_v(i) = v_{i+1} - v_i.
\end{equation}

(3) Apply the sign function to the difference vector ${S}(i) = {\rm {sign } } ({\rm {Diff}}_v)$, where
\begin{equation}
    {\rm {sign }} (x)=\begin{cases}
1,&\text{if   } x > 0;\\
0,&\text{if   } x  = 0;\\
-1,&\text{if   } x  <0.\end{cases}
\end{equation}

(4) Iterate through $S(i)$ from the tail to perform the following:
\begin{equation}
{S}(i)=\begin{cases}
1,\text{if   } {S}(i) = 0 \text{ and } {S}(i+1) \ge 0;\\
-1,\text{if   } {S}(i) = 0 \text{ and }  {S}(i+1) < 0.\end{cases}
\end{equation}

(5) Perform the first-order difference operation on $S(i)$ to obtain $R(i) = {\rm {Diff}}_v(S(i))$.

(6) Traverse the difference vector $R$. If $R(i) = -2$, then $ i + 1$ is crest of the projection vector $V$, and the corresponding crest is $V(i +1)$.  If $R(i) = 2$, then $ i + 1$ is a trough of the projection vector $V$, and the corresponding trough is $V(i +1)$.

\begin{figure}[t!]
    \centering
    \centering
    \subfigure[Wavelet denoised.]{
        \includegraphics[width=3in]{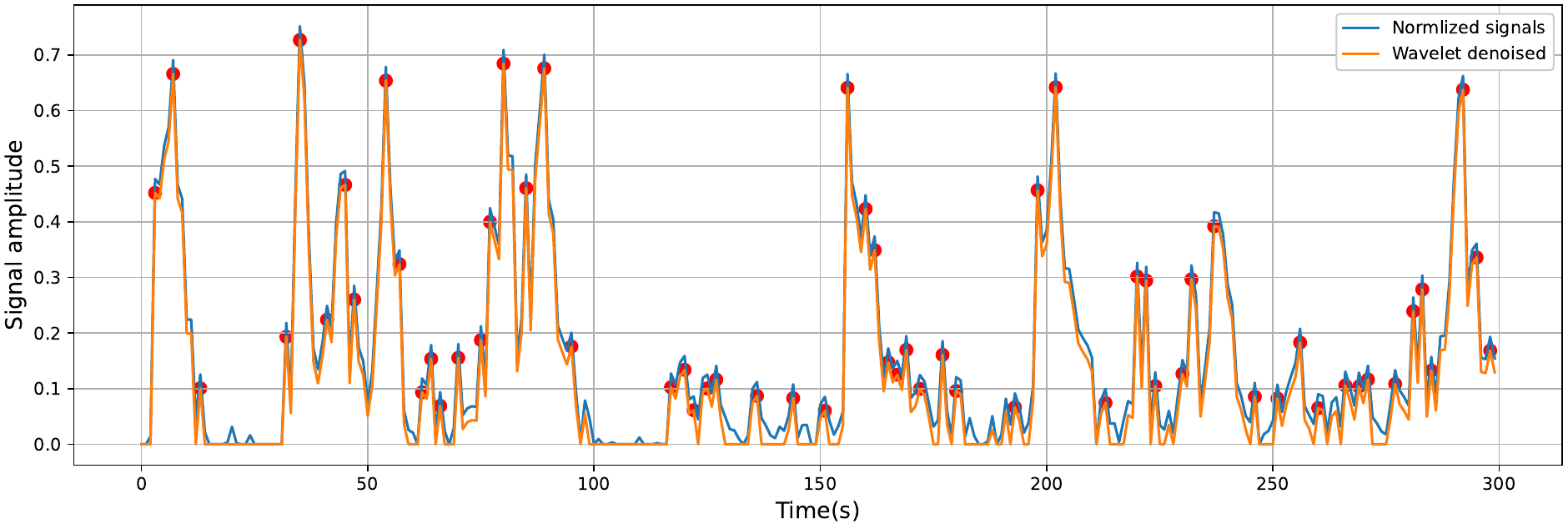}
    }
    \subfigure[Butterworth filtered.]{
        \includegraphics[width=3in]{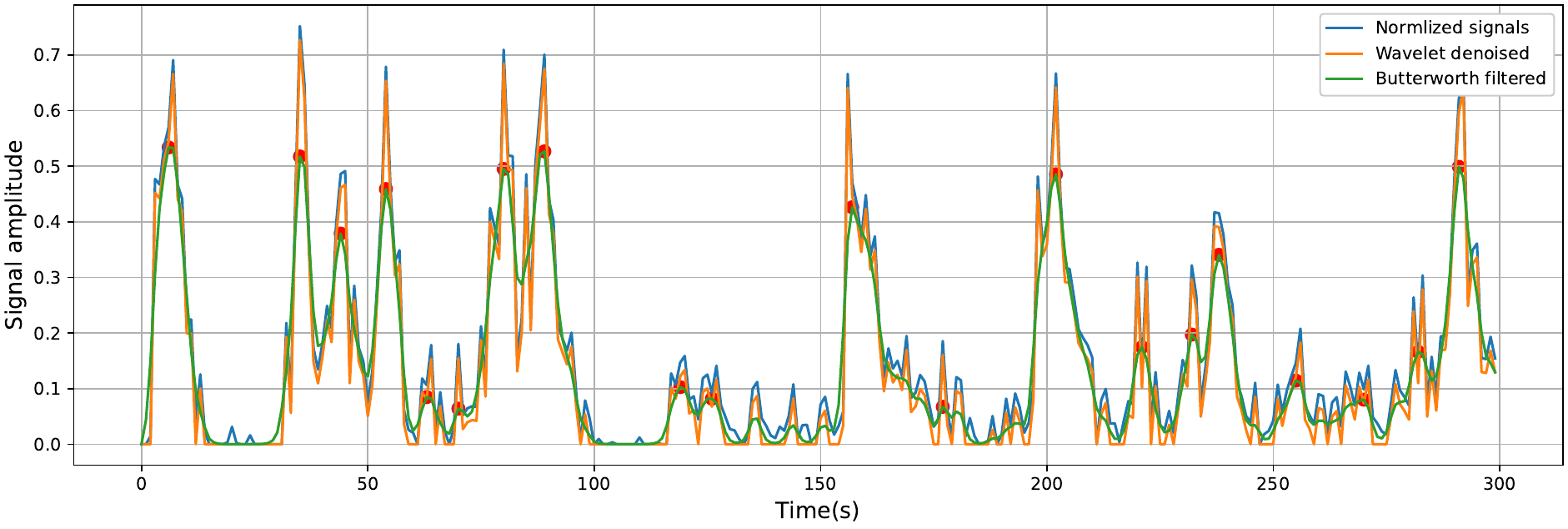}
    }
\caption{Graphical results of peaks location search algorithm in highway scenario. The blue, orange and green lines represent normalized original signals, wavelet denoised signals and Butterworth filtered signals respectively. The red dots represent the detected peaks.}
\label{filtered_signal}
\end{figure}

Here we show the results of peaks location search. As shown in Fig. \ref{filtered_signal}, relatively accurate vehicle position points can be obtained after applying the Butterworth low-pass filter to the waterfall diagrams column-wise and subsequently extracting peaks. Through the visual comparison of Fig. \ref{filtered_signal}(a) and (b), it can be seen that if directly extract peaks from denoised signals, it will produce multiple detection phenomenon, which is inconsistent with the actual situation, so it is necessary to carry out Butterworth low-pass filtering.

\subsection{Vehicle trajectory extraction}

In order to achieve multi-vehicle tracking, we propose a line-by-line matching algorithm based on the vehicle motion model. The vehicle trajectory extraction algorithm flow is shown in Algorithm \ref{algorithm_2}.

\begin{algorithm}[htbp]
    \caption{Trajectory Extraction Algorithm.}
    \label{algorithm_2}
    \LinesNumbered 
    \KwIn{$A = (a_{ij})$: original optical fiber matrix;\ $i = 1, ..., m$, $j = 1, ..., n$, $a_{ij} \in \mathbb{R^{+}}$: selected time and distance.}
  
    \KwOut{$\Sigma^{(S)} = \{(\sigma^{(s)})_{1 \leq s \leq S}\}$: vehicle trajectory point sets.}
  
    Apply Algorithm \ref{algorithm_1} to $A = (a_{ij})$, returning preprocessed optical fiber matrix $D = (d_{ij})$; vehicle counts $S$ and appearance time of each vehicle $k_s$.
  
    $(k_s + \alpha, l_{k_s + \alpha})$ is a pair of time and distance position points of each extracted vehicle trajectory\;
    $l_{k_s + \alpha}\in \mathbb{N^{+}}$ is distance position point corresponding to time position point $k_s + \alpha$ in $D = (d_{ij})$\;
    $\Sigma^{(S)} = \{(\sigma^{(s)})_{1 \leq s \leq S}: \sigma^{(s)} = (k_s + \alpha, l_{k_s + \alpha}), \alpha \in \mathbb{N}$\}\;
    
    Select the initial velocity interval $v_{\min}^{(1)}$ to $v_{\max}^{(1)}$\;
    Calculate the initial distance interval  $x_{\min}^{(1)}$ to $x_{\max}^{(1)}$\;

    \ForEach{$s = 1$ to $S$; $\sigma^{(s)} \in \Sigma^{(S)}$}{
      Add the first pair $(k_s, l_{k_s})$ to vehicle trajectory point sets $\sigma^{(s)}$, where $l_{k_s} = 1$\;
      $l_{k_s + 1}$ $\gets$ ColIndex(Max ($D_{k_s + 1, l_{k_s}+ x_{\min}^{(1)}: l_{k_s} + x_{\max}^{(1)}}$))\;
       \eIf{$k_s + 1 < m$ $\rm and$ $l_{k_s + 1} < n$}{
          Add pair $(k_s + 1, l_{k_s + 1})$ to each vehicle trajectory point set $\sigma^{(s)}$\;
       }{
          break.
       }
      } 

    \ForEach{$s = 1$ to $S$; $\sigma^{(s)} \in \Sigma^{(S)}$}{
      \For{$\alpha \in \mathbb{N^{+}}$}{
        $y^{(s)}$ = PolynomialFitting($\sigma^{(s)}$)\;
        $v^{(s)}$ = Slope($y^{(s)}$)\;
        Select the right level of confidence $\rm cof$\;
        Calculate velocity interval $(1 + \rm cof)$$v^{(s)}$ to $(1 + \rm cof)$$v^{(s)}$\;
        Calculate distance interval $x_{\min}^{(s)}$ to $x_{\max}^{(s)}$\;
        $l_{k_s + \alpha}$ $\gets$ ColIndex(Max ($D_{k_s + \alpha, l_{k_s + \alpha - 1}+ x_{\min}^{(s)}: l_{k_s + \alpha - 1} + x_{\max}^{(s)}}$))\;
       \eIf{$k_s + \alpha < m$ $\rm and$ $l_{k_s + \alpha} < n$}{
          Add pair $(k_s + \alpha, l_{k_s + \alpha})$ to each vehicle trajectory point set $\sigma^{(s)}$\;
       }{
          break.
       }
      } 
      }
    \textbf{Return:} Vehicle trajectory point sets $\Sigma^{(S)}$.
\end{algorithm}

Different from previous work \cite{van2022deep, wiesmeyr2021distributed, catalano2021automatic, wang2021vehicle}, complete trajectories of each vehicle in the fixed-size optical fiber diagrams waterfall diagram is regarded as straight lines, that is, the vehicle moves uniformly throughout the whole process. Our algorithm uses this feature locally rather than globally. In each time minimum interval, we view the vehicle as moving in a straight line at uniform speed.

According to Newton's first law, vehicles can be regarded as moving at a uniform velocity in the absence of lane changes, emergency stops, or overtaking maneuvers. The space-time information reflected by optical fiber within a tiny time unit exhibits a straight-line pattern. This characteristic aligns well with a uniform linear motion model represented as:
\begin{equation}
x = vt + x_0,
\end{equation}
where $v$ is velocity, $x_0$ is the initial position where the vehicle appears, and $x$ is the vehicle's position at time $t$. 

\begin{figure}[t!]
    \centering
    \includegraphics[width=3in]{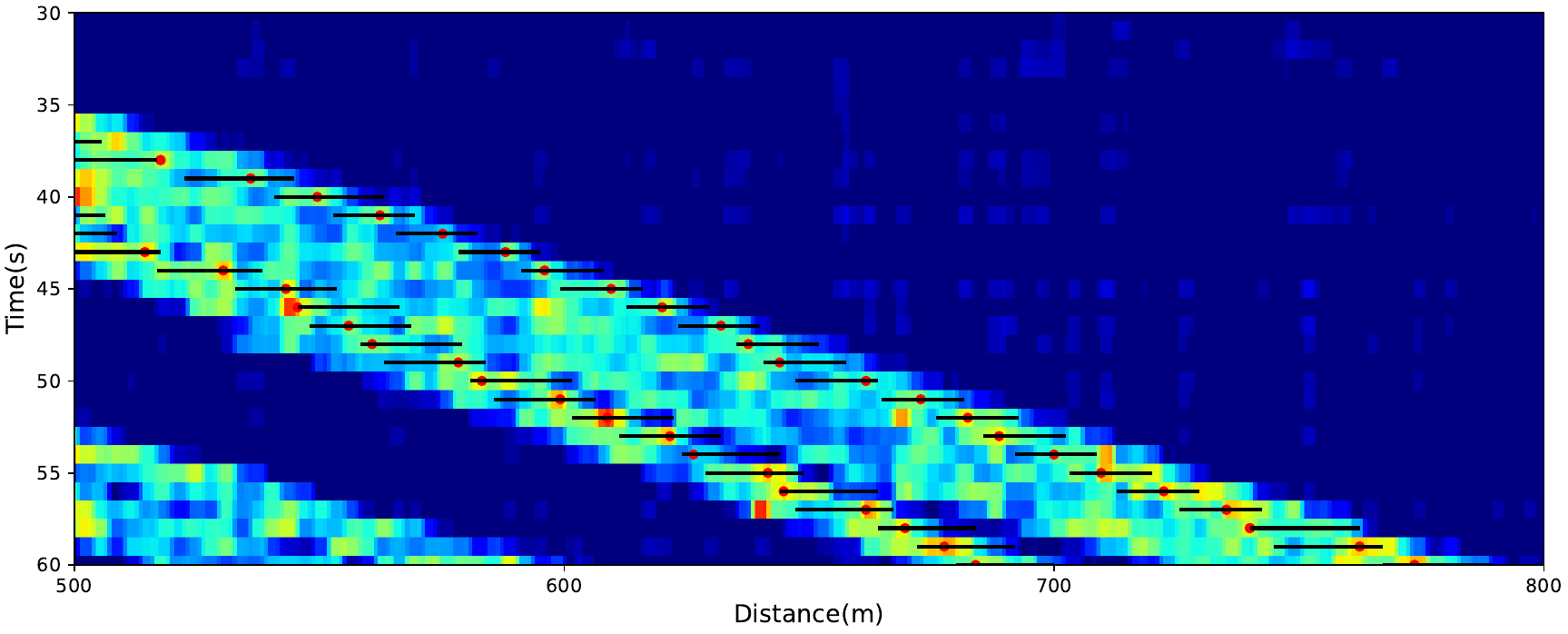}
\caption{Trajectory extraction algorithm. The black line segments represent the distance range and the red dots represent trajectory key points.}
\label{prior_improved}
\end{figure}

Therefore, after determining the starting position via vehicle position detection, knowing the vehicle's velocity enables the determination of its position along its trajectory. However, in the case of vehicle congestion, there is a risk of multiple vehicles overlapping. To address this challenge, we expand the speed range of each vehicle from a static line segment to a dynamic line segment. The selection of this dynamic line segment is determined by the slope of the line fitted using historical position and velocity, incorporating a certain degree of confidence. Fig. \ref{prior_improved} illustrates the dynamic line segments when vehicle congested, facilitating obtaining key points in vehicle trajectories through line-by-line matching.

Given the key points of vehicle trajectories, we can reconstruct the complete vehicle trajectories and subsequently calculate the instantaneous speed of the vehicles. Furthermore, in scenarios where most short-distance vehicles on the highway maintain a consistent speed, we can further employ linear fitting to derive the average speed of these vehicles. To achieve this, we utilize a straightforward curve-fitting technique known as polynomial fitting, defined as follows: 
\begin{equation}
y(x, \omega) = \omega_0 + \omega_1x +\omega_2x+\ldots+w_Mx^M = \sum_{j = 0}^{M}\omega_jx^j,
\end{equation}
where $m$ is the order of the polynomial, $\omega = (\omega_0,...  \omega_M)^T$ represents the polynomial coefficients as a vector. Our objective is to obtain the coefficients of the fitting polynomial by minimizing the error between the fitting function and the data points in the training set: observations of $(x,t)$, $x = (x_0,...  x_N)^T$ and $t = (t_0,...  t_N)^T$. We employ the mean square error function, expressed as:
\begin{equation}
E(\omega) = \frac{1}{2}\sum_{n = 1}^{N}(y(x_n, \omega) - t_n)^2,
\end{equation}
where the error function is quadratic with respect to the coefficient $\omega$ and has a non-negative value. When the error function is minimized, the unique solution $w^*$ along with the fitting polynomial $y(x, w^*)$ can be obtained. 

\begin{figure}[htbp]
    \centering
    \subfigure[Trajectory key points.]{
        \includegraphics[width=3in]{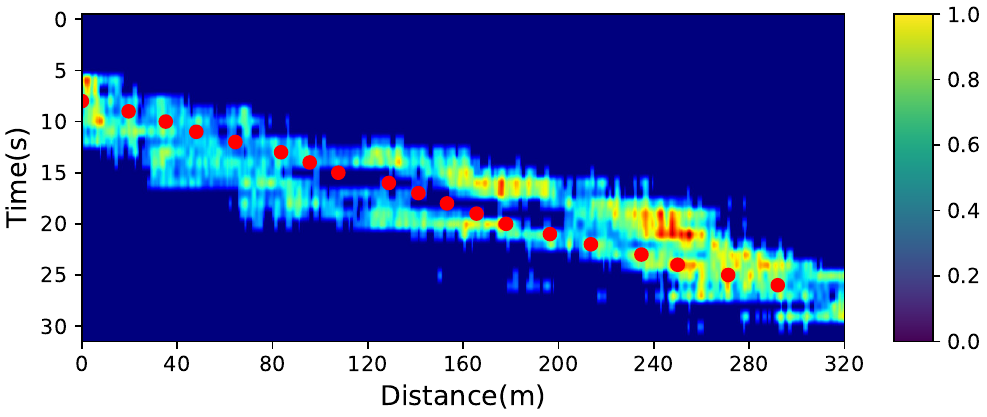}
    }
    \subfigure[Trajectory fitting results.]{
        \includegraphics[width=3in]{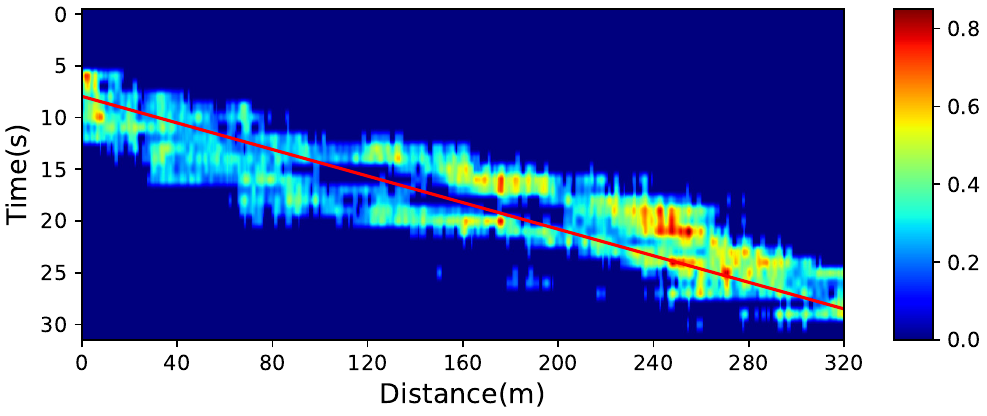}
    }
\caption{Trajectory extraction results of a large truck. Even if the DAS system detects the truck as two trajectories in the middle sections, our algorithm can still recover the complete trajectory.}
\label{track}
\end{figure}

Here we illustrate the effectiveness of our proposed tracking method by taking a large truck as an example. Because the truck is too long, two separate trajectories are wrongly detected in the middle sections, as illustrated in Fig. \ref{track} (a). However, our algorithm will detect the key points within the given speed range every second after determining the truck's entry position. Even if some of the detected key points do not fall on one of the two separate trajectories, it can still reflect the approximate route, which is consistent with the actual situation. Based on key points for a single vehicle, we can deduce the vehicle's approximate uniform-speed trajectory through polynomial fitting, as shown in Fig. \ref{track} (b). 

\section{Experiments}
\label{Experiments}

We introduce our self-established highway and tunnel datasets in \ref{Self-established Datasets}. Standard indices regarding experiments in traffic monitoring scenario are illustrated in \ref{Metrics}. Besides, based on these standard indices, qualitative and quantitative analysis of our comparative experiments are conducted in tunnel and city scenarios in \ref{analysis}.

\subsection{Datasets}
\label{Self-established Datasets}
In this section, we conducted data acquisition and analysis in real highway scenario, with a particular focus on a tunnel scenario. The DAS systems are laid 5 to 10cm outside the driveway, with a depth of about 5cm underground. The specific DAS recording parameters are shown in Table \ref{configuration}. As our DAS systems were laid on only one side of the road, the target areas for this paper are explicitly stated. The algorithm proposed aims to extract vehicle trajectories within these target areas. The Jingjintang expressway dataset is used to verify the validity of DAS systems for large-scale traffic monitoring. The Jiurui tunnel dataset and a public city dataset in \cite{van2022deep} are used to verify the effectiveness of the proposed algorithm in extracting optical fiber signals. As the description of public city dataset can be found in \cite{van2022deep}, here we only describe our self-established two datasets.

\begin{table}[!t]
    \caption{DAS recording parameters.}
    \label{configuration}
    \centering
    \renewcommand\arraystretch{1.2}
    \begin{tabular}{ccc}
    \hline
    Parameter & Jingjintang expressway dataset & Jiurui tunnel dataset\\
    \hline
        Pulse repetition frequency & 4000Hz & 4000Hz \\
        Gauge length & 300m & 5000m \\
        Sample spacing & 40cm & 80cm \\
        ADA sampling frenquency & 4Gbps & 4Gbps \\
        \hline
    \end{tabular}
\end{table}

\textbf{Target area in Jingjintang expressway dataset:} Located in Junliang City Test Site, Dongli District, Tianjin, China, Jingjintang expressway is a dual carriageway with three lanes in each direction. Our target area encompasses three lanes with a shoulder on the right, excluding the opposite lanes.

\textbf{Target area in Jiurui tunnel dataset:} The Jiujiang-Ruichang expressway, situated in Jiujiang City, Jiangxi Province, China, represents a real tunnel scenario with a single carriageway featuring two lanes. Our target area focuses on the right lane, disregarding the left lane.

To assess the effectiveness of the optical fiber equipment, cameras were installed on the same right side of the road for data comparison in Jingjintang expressway dataset. Our camera data is used here solely to evaluate the reliability of data collected by optical fiber. The algorithm proposed in this paper exclusively relies on optical fiber data and is independent of camera data.

\subsubsection{Jingjintang Expressway Dataset}
\label{Jingjintang Expressway Dataset}

\begin{figure}[htbp]
    \centering  
    \includegraphics[width=0.49\textwidth]{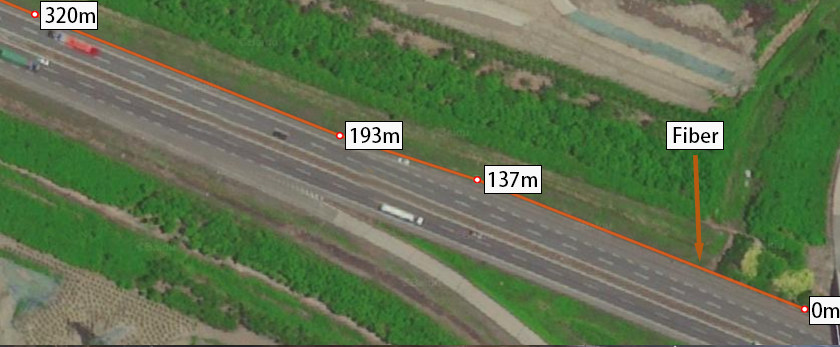}
    \caption{Overhead view of the Jingjintang expressway, and the optical fiber with a length of 320m is laid on one side of the road.}
    \label{highway_view}
\end{figure}

\begin{figure}[!t]
    \centering
    \includegraphics[width=2.5in]{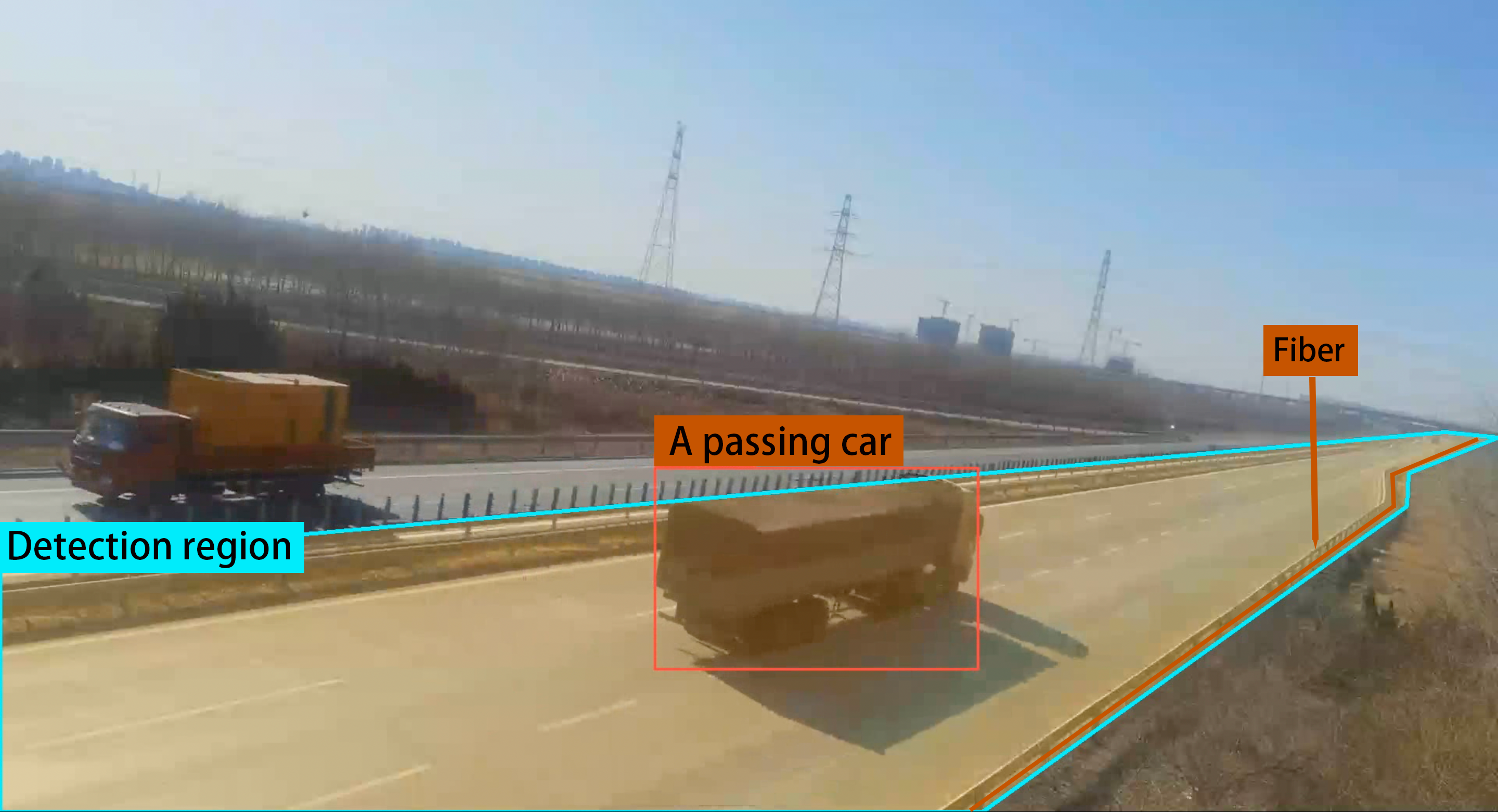}
    \caption{Camera detection of the target area, including 3 lanes and a shoulder.}
    \label{camera}
\end{figure}

We selected a 320-meter section of Jingjintang expressway for our study and optical fiber sampling interval is 0.4m. The aerial view of detected area is shown in Fig.\ref{highway_view}. Simultaneously, a camera mounted on the road captures movement information of the vehicles. Positioned at the starting point of the optical fiber laying side, the camera primarily monitors vehicles in the target area, as depicted in Fig. \ref{camera}.

\begin{figure}[t!]
    \centering
    \subfigure[3 trucks.]{
    \includegraphics[width=1in]{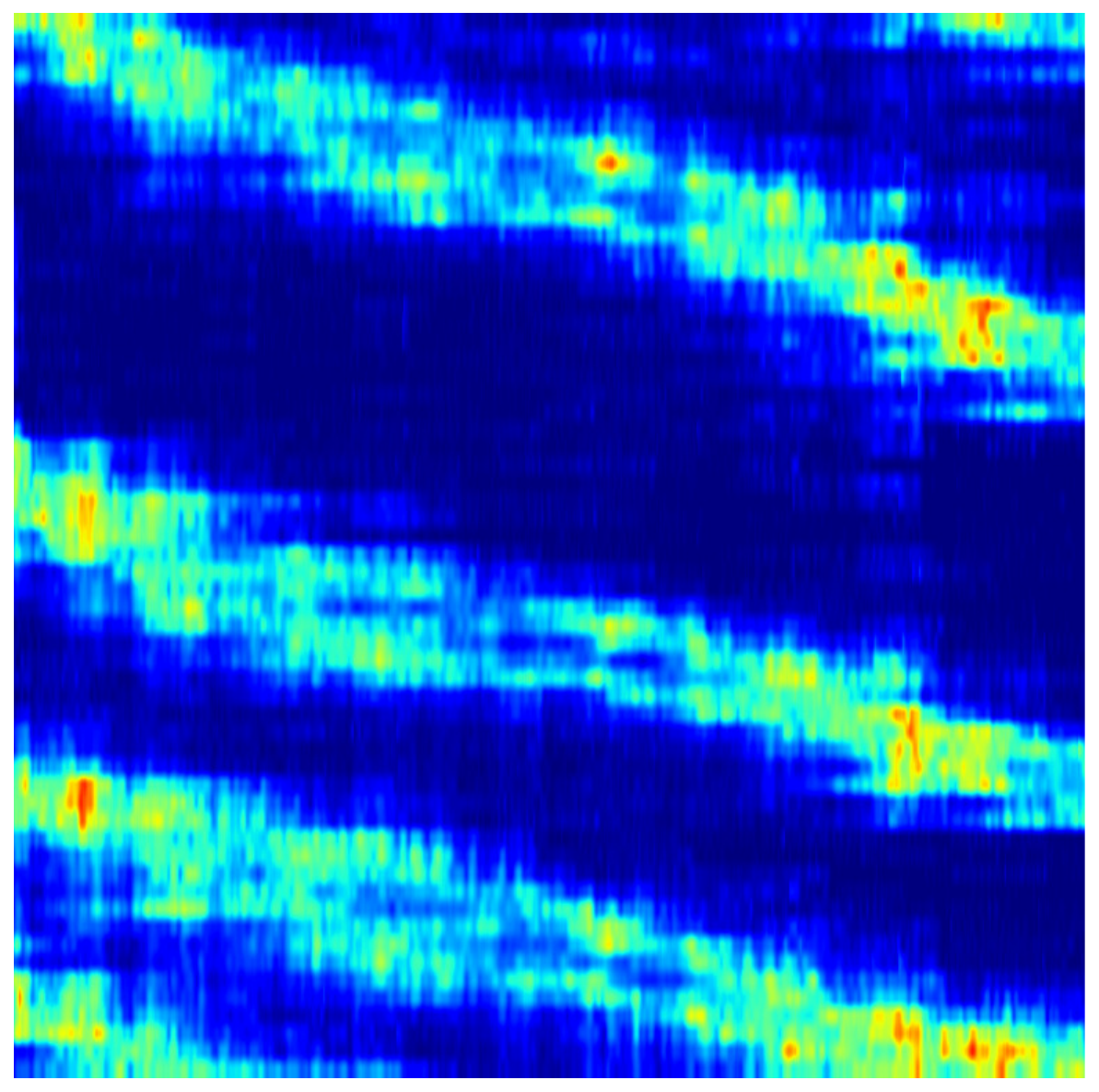}
    }
    \subfigure[3 cars.]{
    \includegraphics[width=1in]{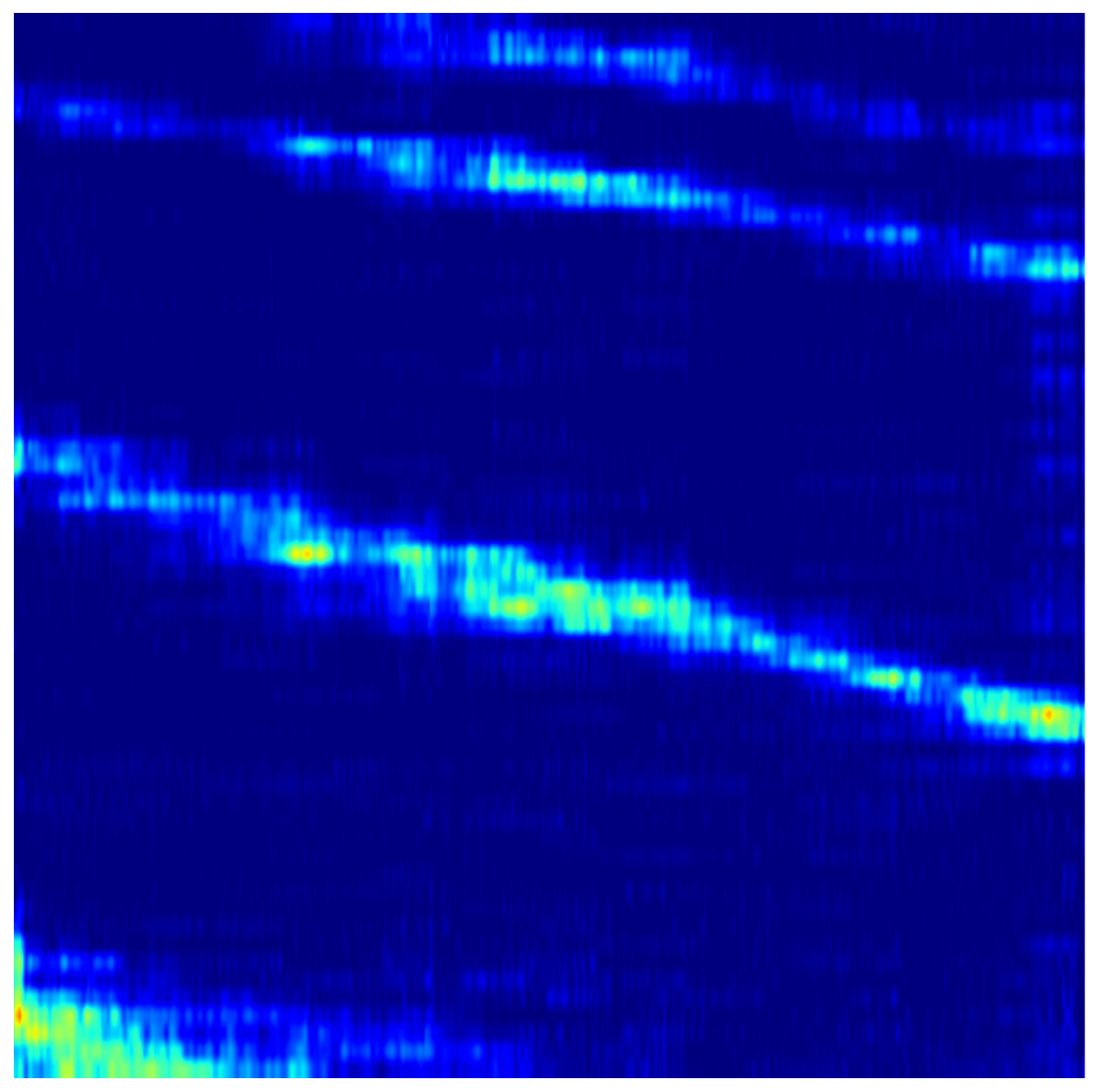}
    }
    \subfigure[1 truck and 4 cars]{
    \includegraphics[width=1in]{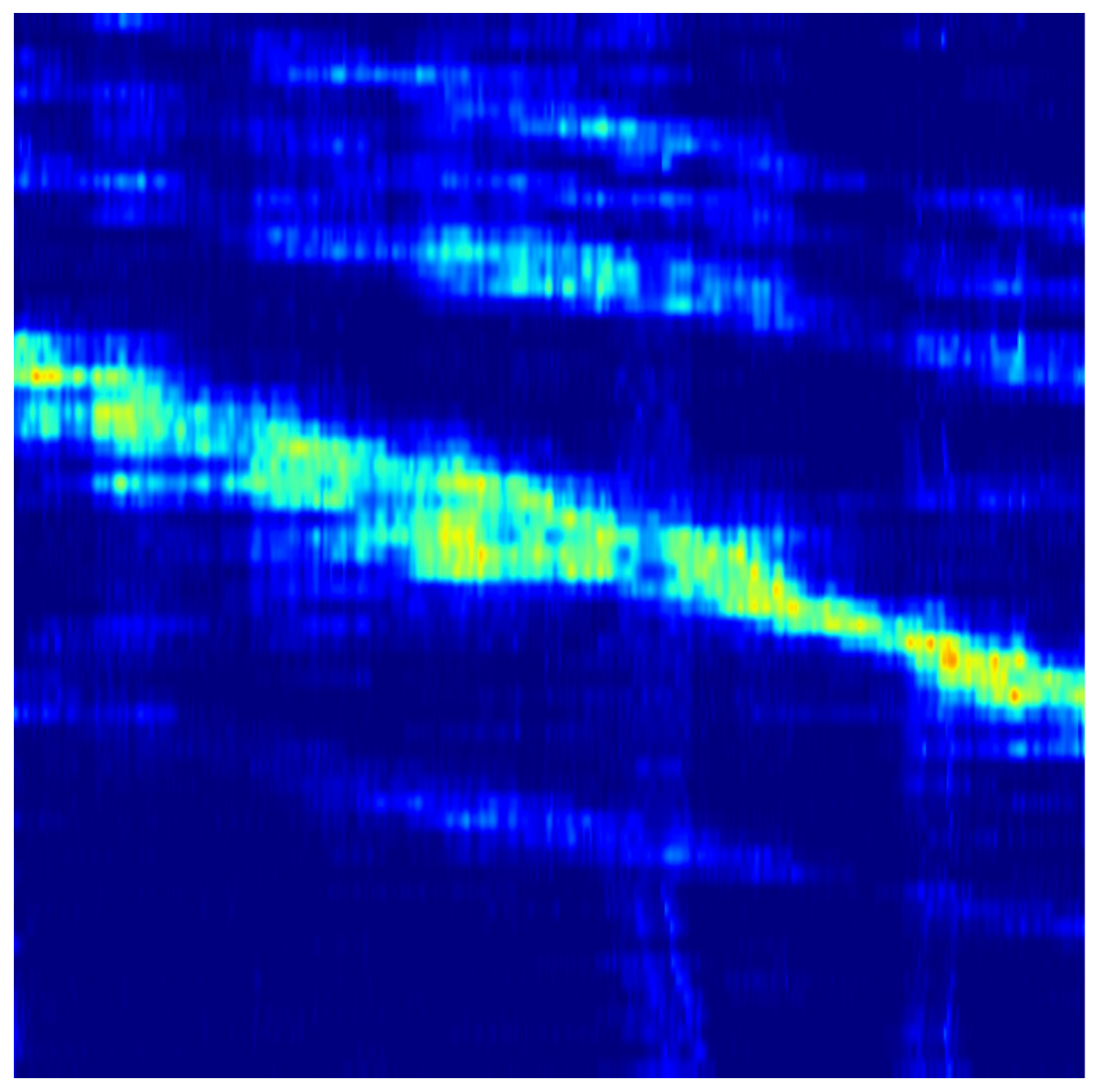}
    }
\caption{Simple cases in Jingjintang expressway dataset. Vehicles pass in turn, without interference with each other, and without interference from the opposite lanes.}
\label{simple}
\end{figure}

\begin{figure}[t!]
    \centering
    \subfigure[Traffic flow with cars and trucks.]{
        \includegraphics[width=1in]{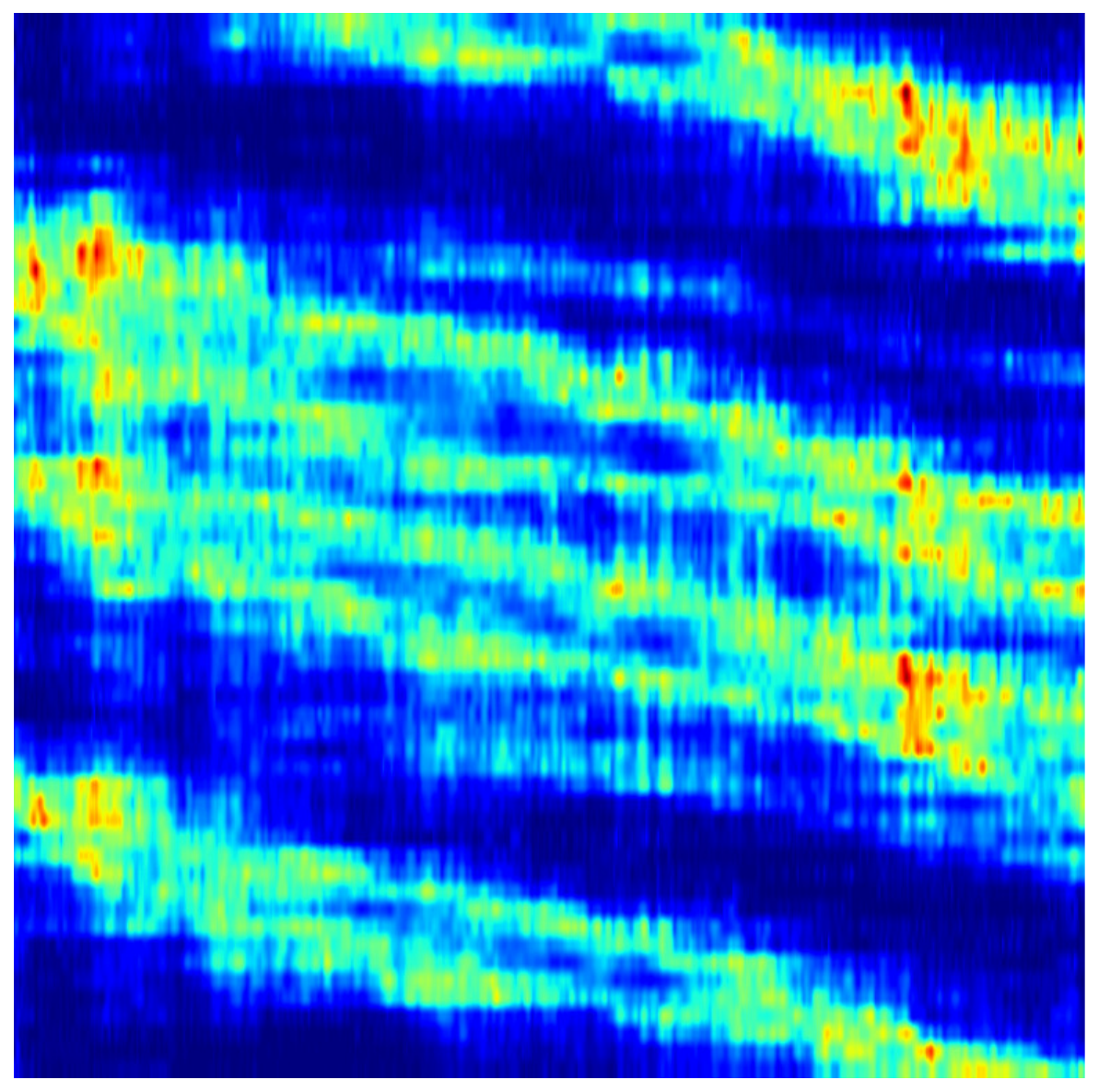}
        \includegraphics[width=1in]{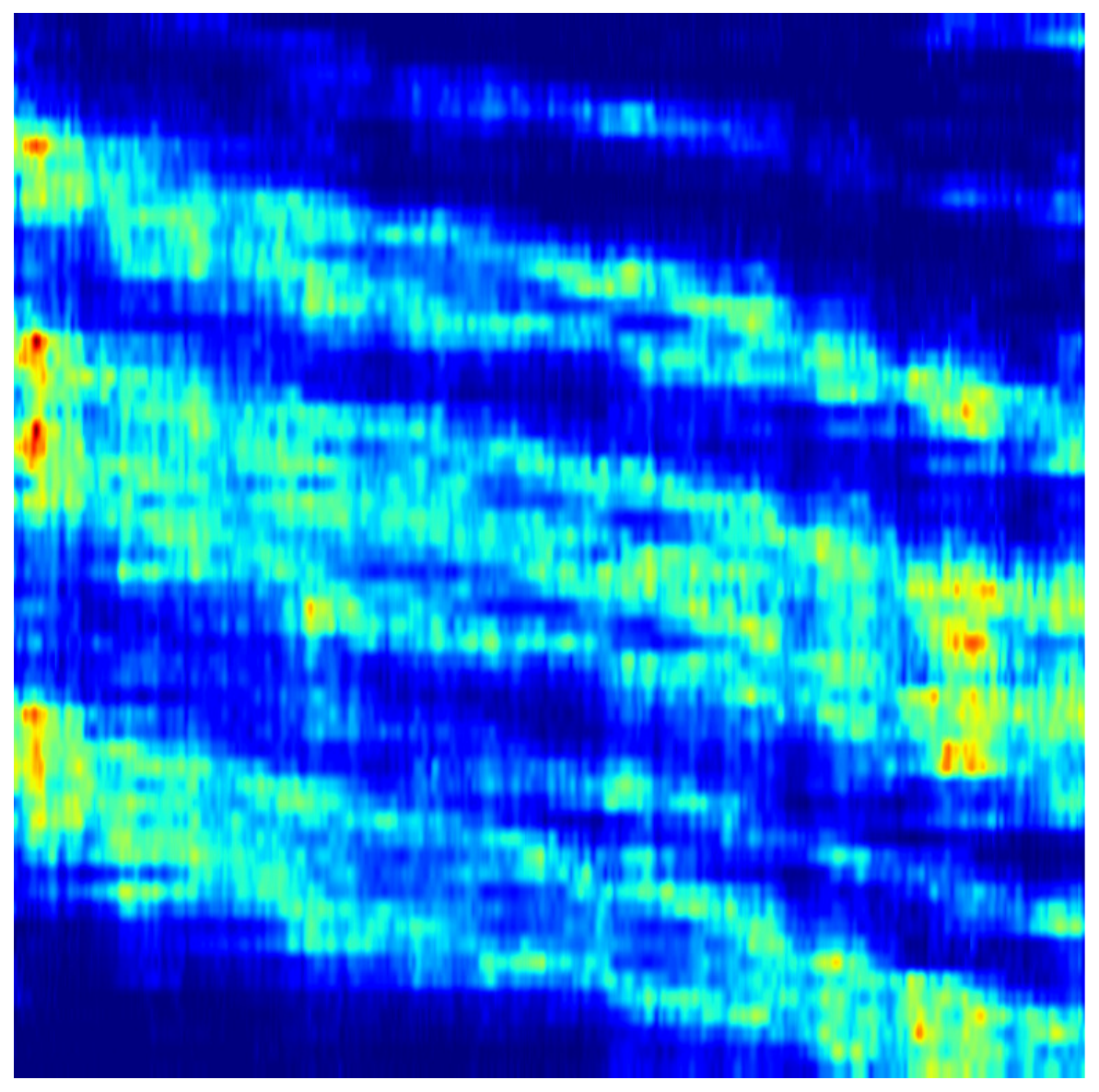}
        \includegraphics[width=1in]{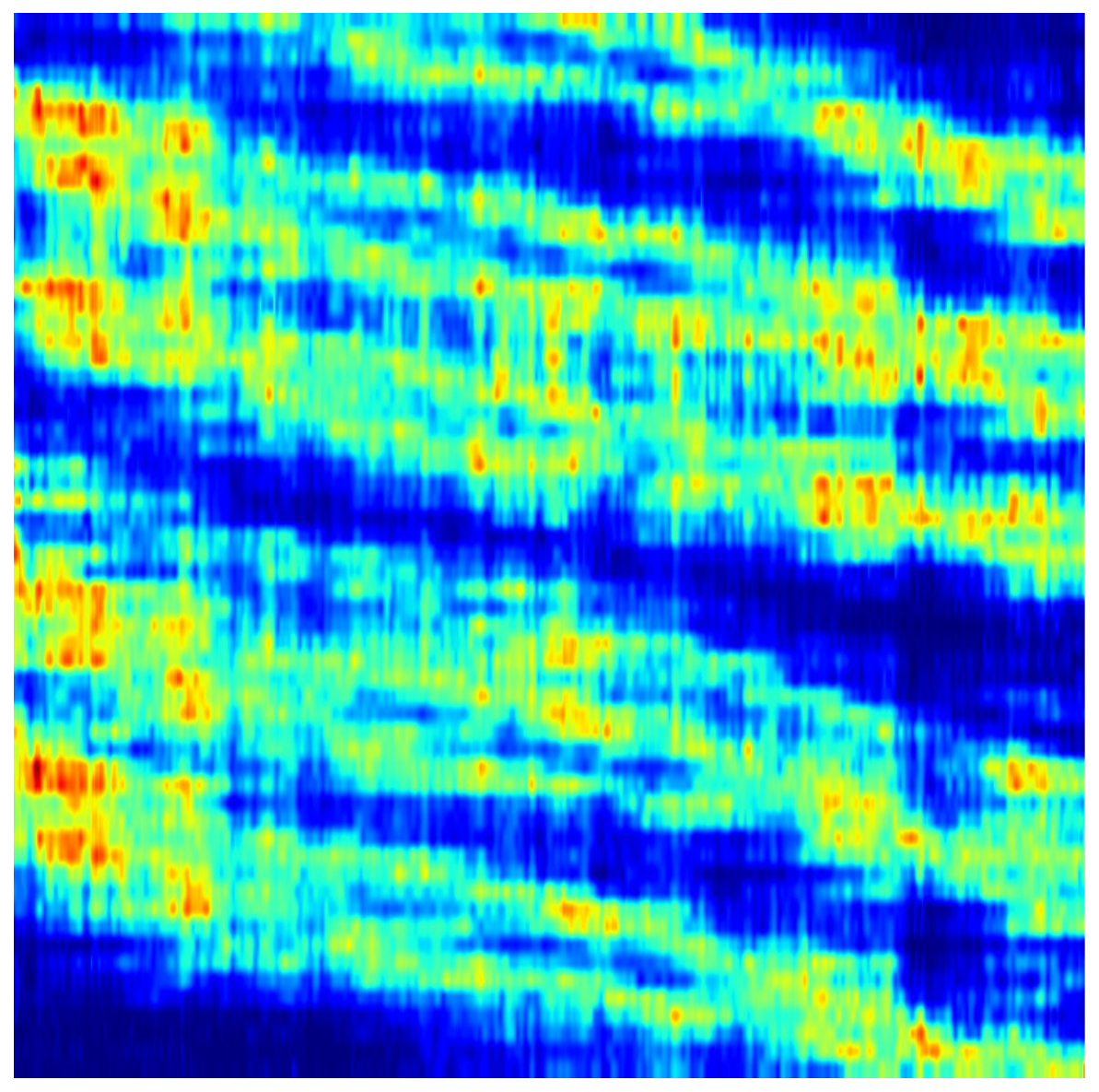}
    }\hfill
    
    \subfigure[Multi-vehicle parallel.]{
    \includegraphics[width=1in]{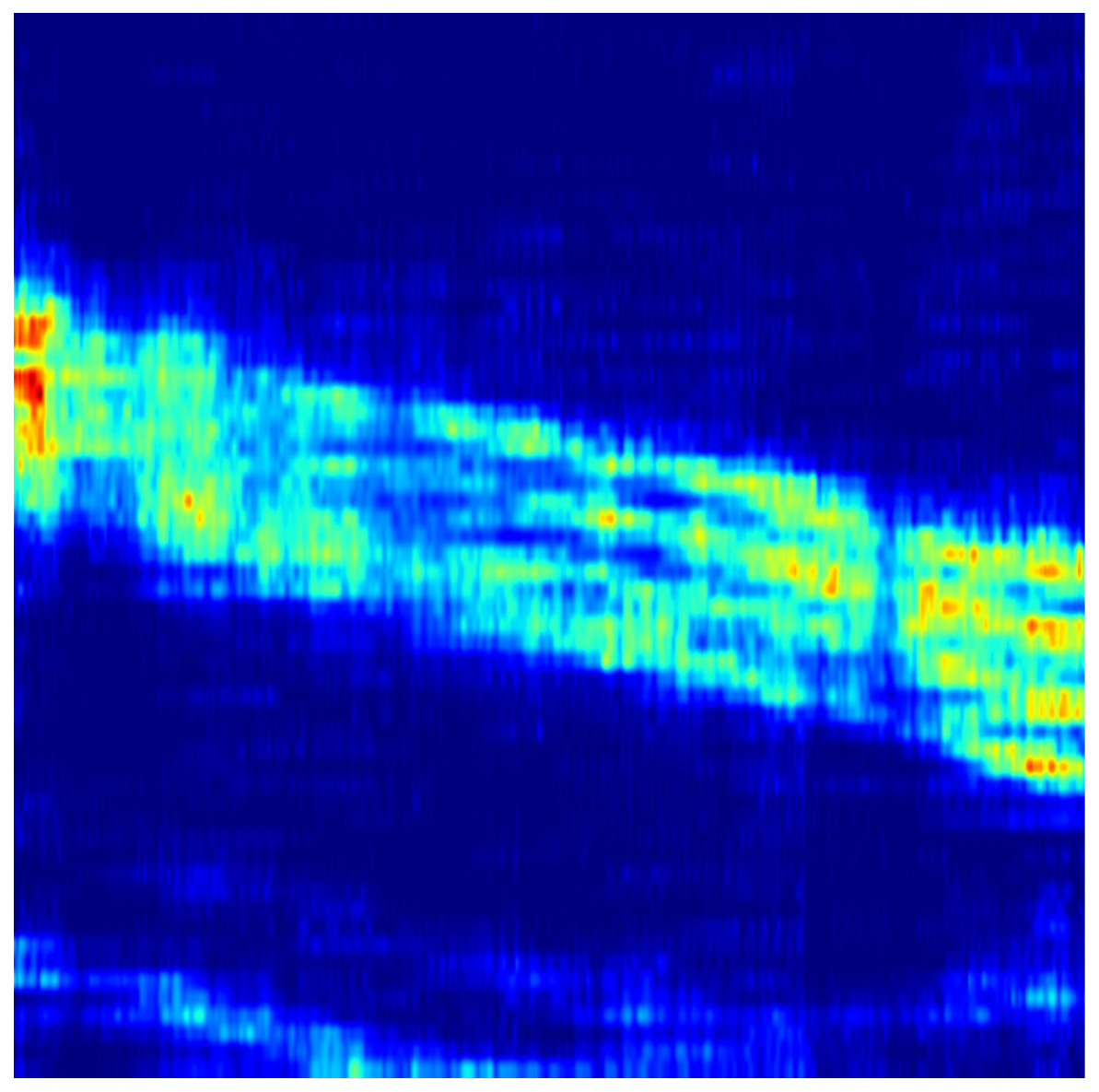}
    \includegraphics[width=1in]{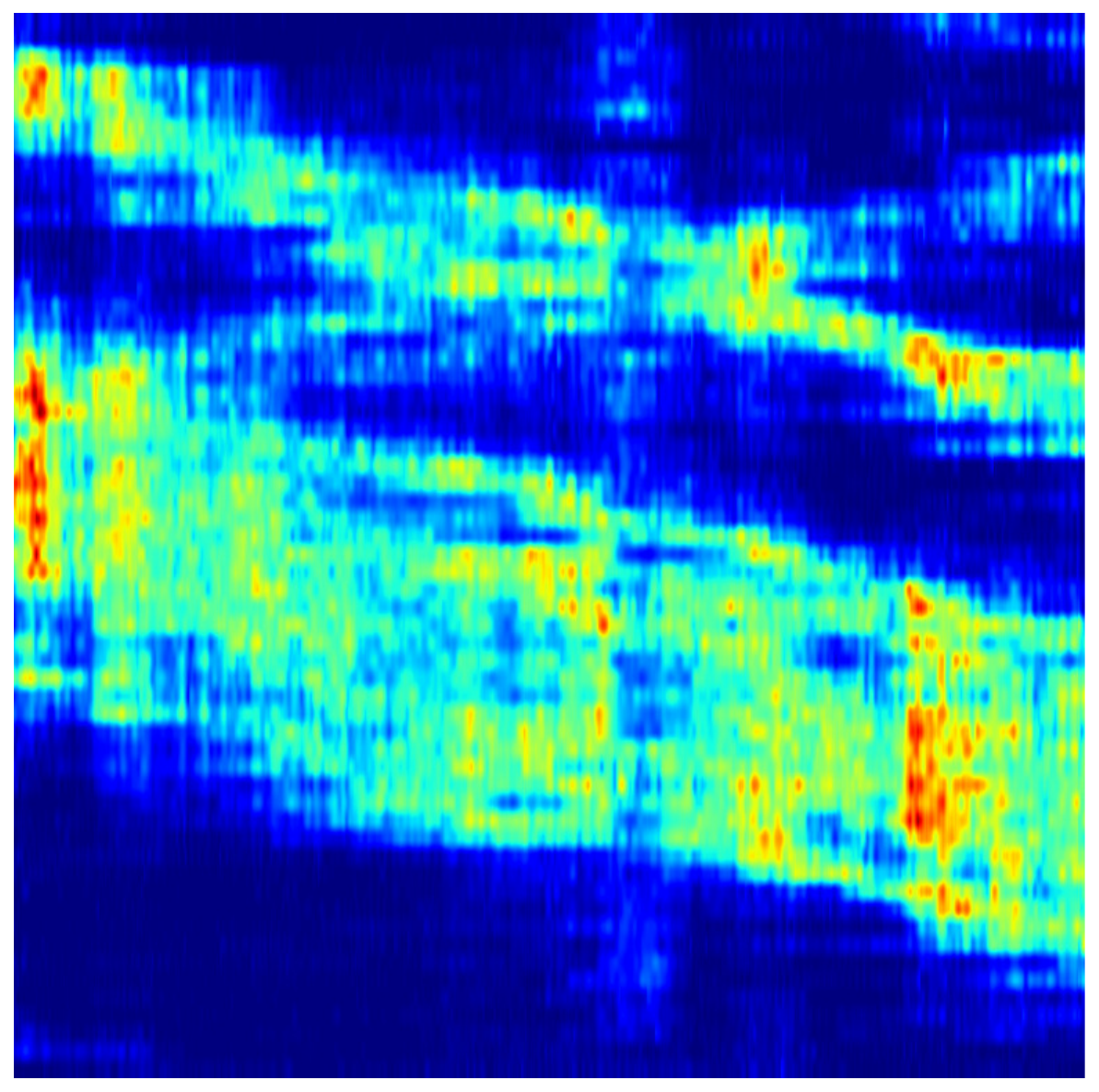}
    \includegraphics[width=1in]{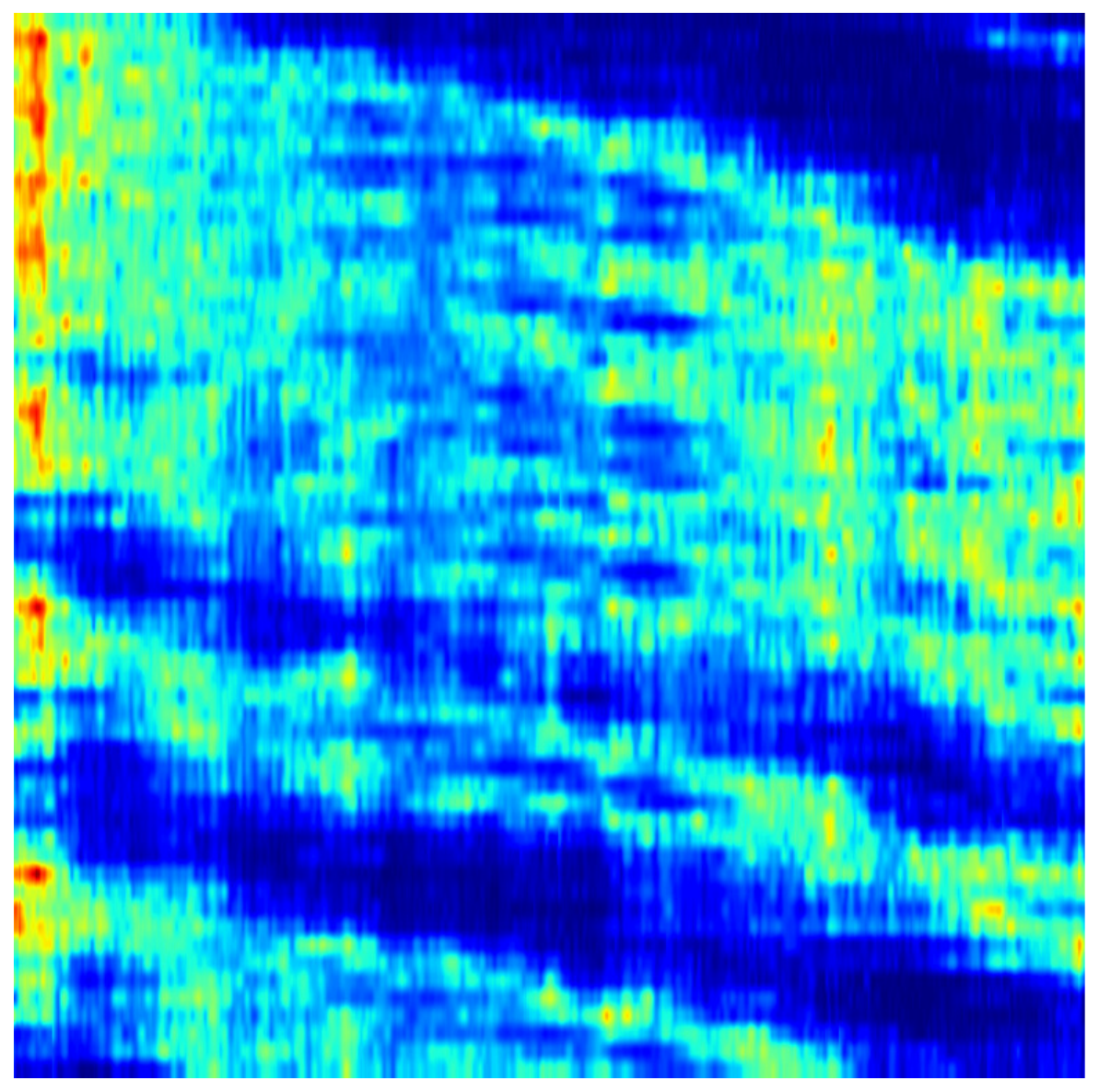}
    }\hfill

    \subfigure[Opposite interference.]{
    \includegraphics[width=1in]{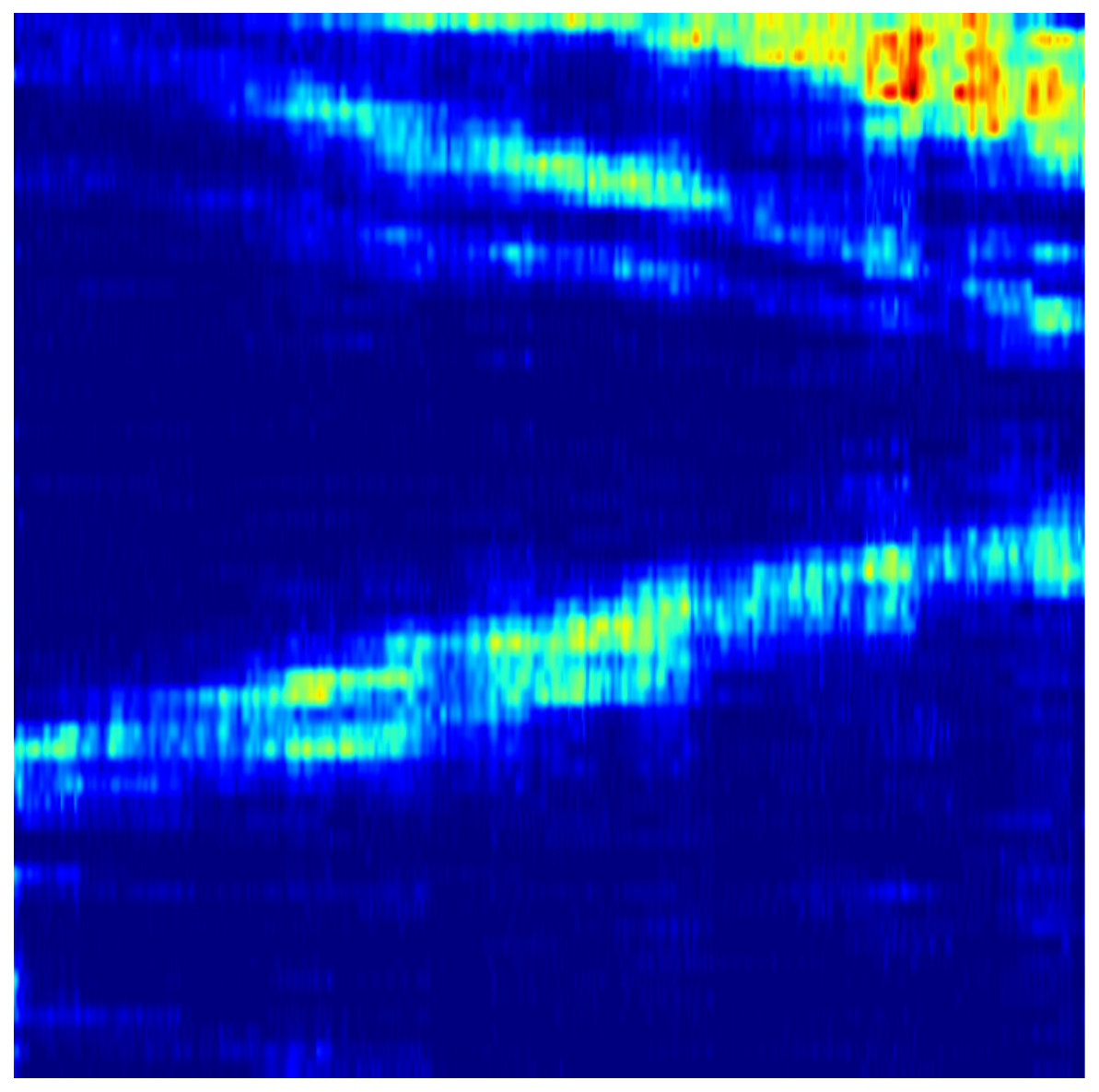}
    \includegraphics[width=1in]{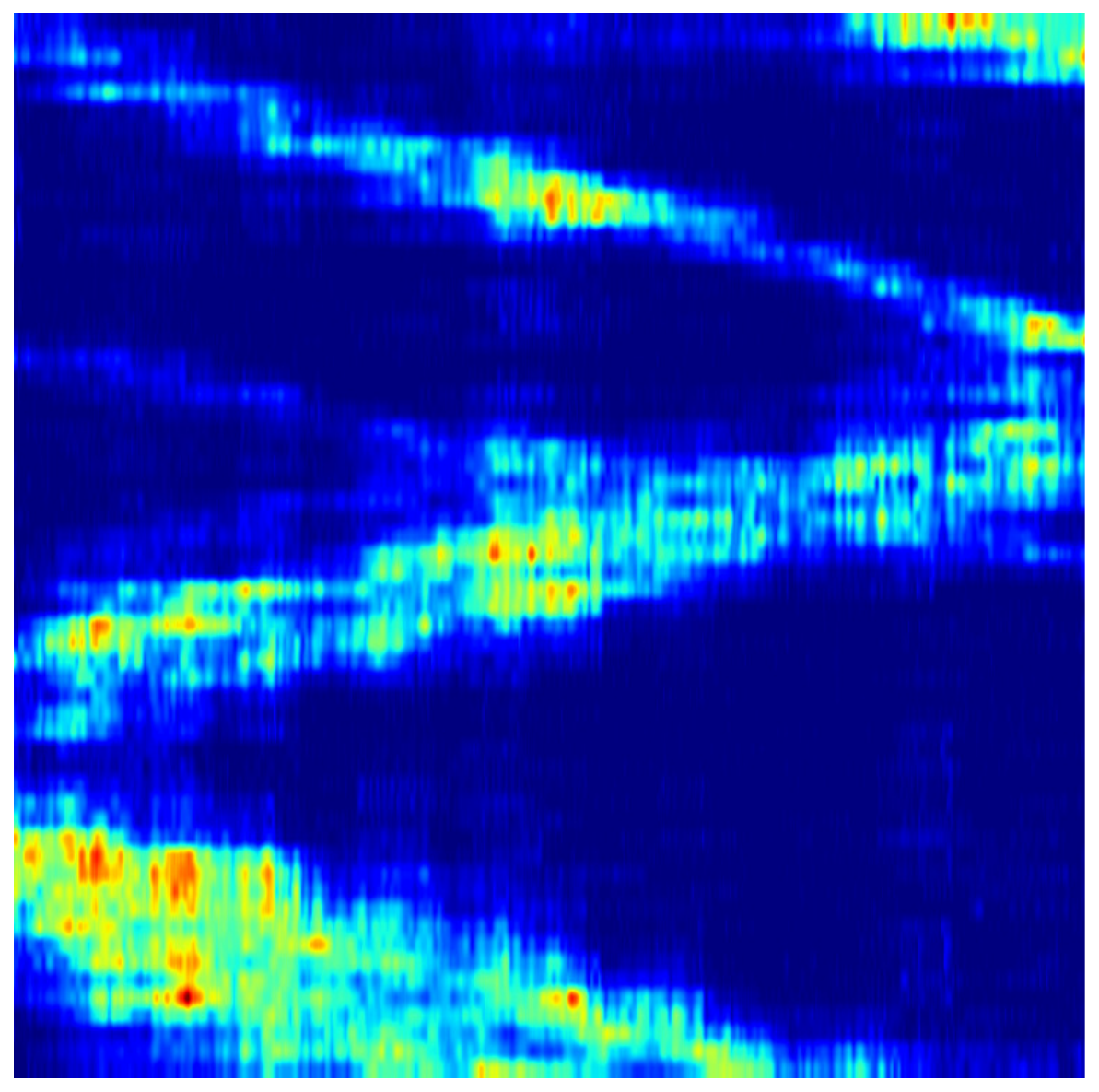}
    \includegraphics[width=1in]{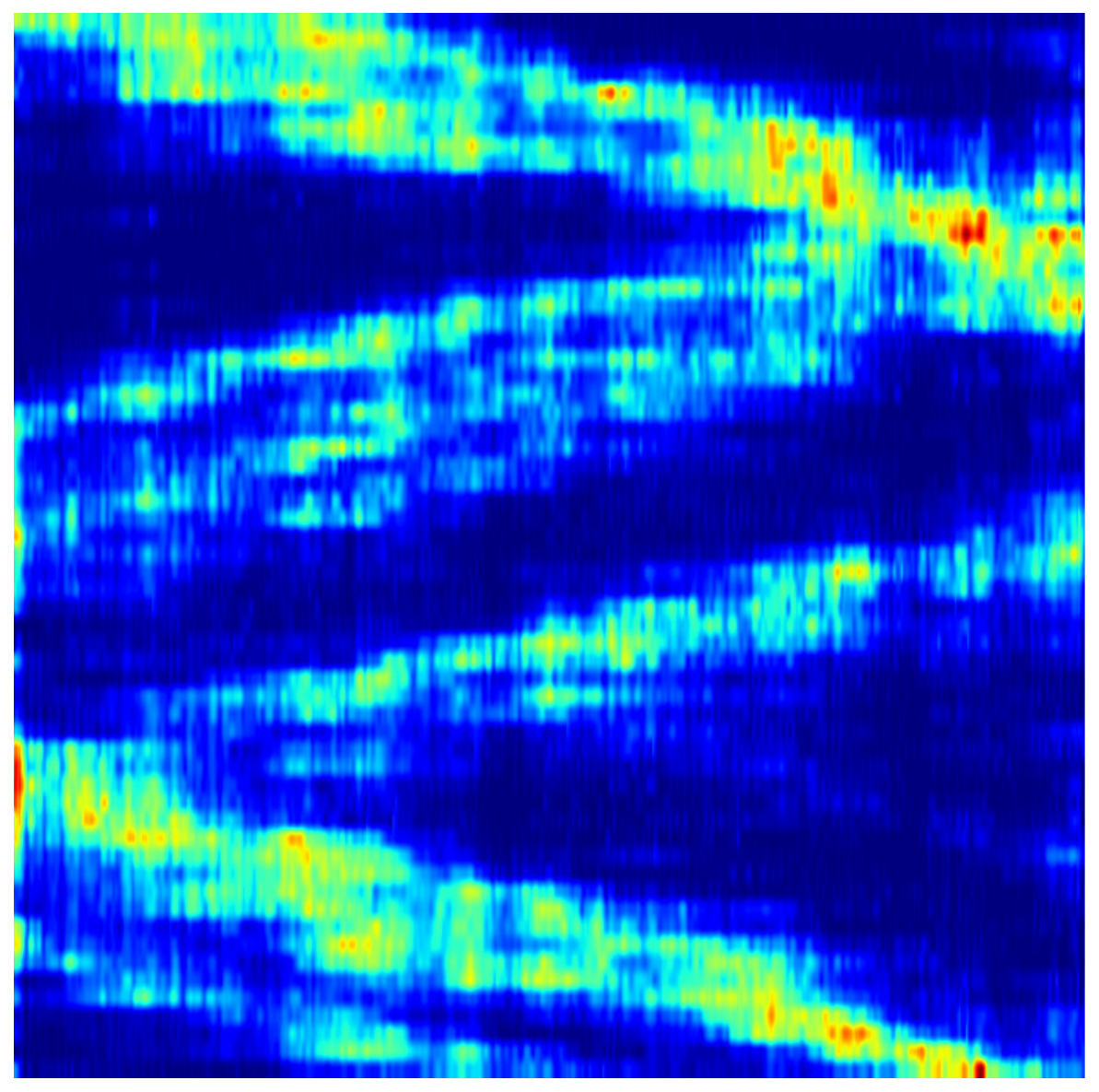}
    }
\caption{Complex cases in Jingjintang expressway dataset. (a) Large trucks significantly overlap neighboring small cars. (b) Vehicles pass nearly simultaneously in different lanes. (c) Vehicles from the opposite lane causing interference with vehicles in the target area.}
\label{complex}
\end{figure}

When a vehicle traverses the target area, the DAS systems record corresponding vibration signals. To present the vehicles' trajectory more intuitively, we utilize waterfall diagrams to plot signal amplitudes, as illustrated in Fig. \ref{simple} and Fig. \ref{complex}. 

In the waterfall diagrams, the light blue color represents background noise, while prominently colored lines indicate vehicles. The fiber stripe extending from the top left to the bottom right corresponds to vehicles within the target area, and the stripe extending from the top right to the bottom left represents vehicles on the opposite side. Notably, the vehicle's trajectory is depicted as an approximately straight line within the diagram. This is attributed to the fact that most vehicles on the highway maintain high and consistent velocities when traveling in lanes with unchanged speeds. In this representation, the intercept of the straight line signifies the starting position of the vehicle, and the absolute value of the slope represents its velocity.

Optical fiber waterfall diagrams excel at representing real-time vehicle location data, especially in simple cases where vehicles pass in sequence without interfering with each other, as illustrated in Fig. \ref{simple}. But during congested period, highway scenario can be more complicated and exist certain inevitable challenges.

(1) Owing to the high sensitivity and rapid response characteristics of optical fiber, large trucks are prone to generating larger signal amplitudes. In instances where a car passes concurrently close, the car's signals may become entirely submerged beneath the truck's signals. As shown in Fig. \ref{complex}(a), it is almost impossible to detect car signals which merge with truck signals.

(2) Considering the relatively large number of vehicles on highways, each vehicle generates substantial signals. When multiple vehicles traverse the highway consecutively, they tend to interfere with each other. As depicted in Fig. \ref{complex}(b), multiple vehicles overlap with each other and are difficult to distinguish within this cluster.

(3) While the signals from vehicles in the opposite lane are not as clear as those within the target area, signals from trucks in the opposing lane can still interfere with the traffic signal data we aim to capture, as shown in Fig. \ref{complex}(c).

\begin{figure}[t!]
    \centering
    \subfigure[Trucks.]{
        \centering
        \includegraphics[width=3in]{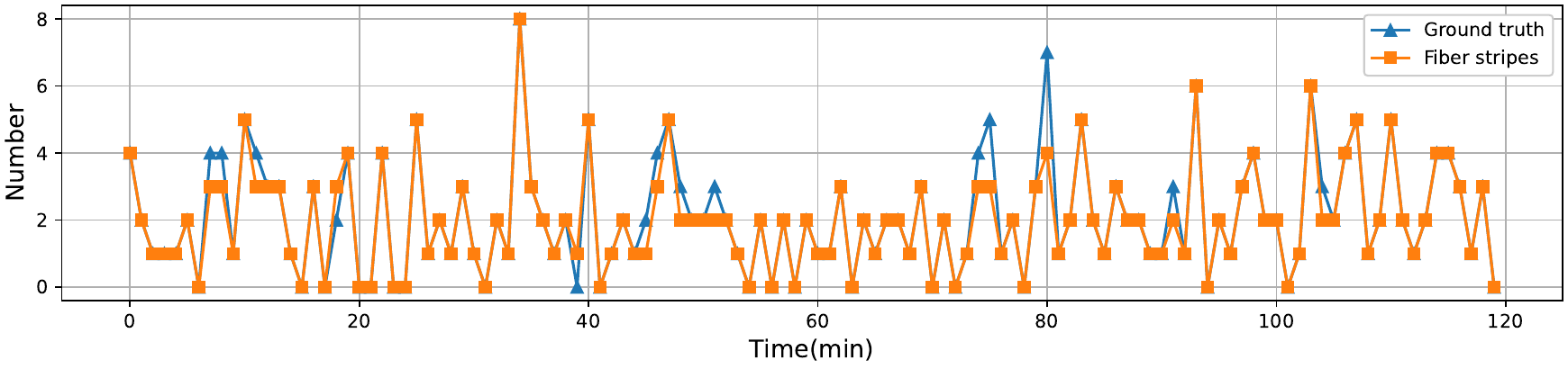}
    }
    \subfigure[Cars.]{
        \centering
        \includegraphics[width=3in]{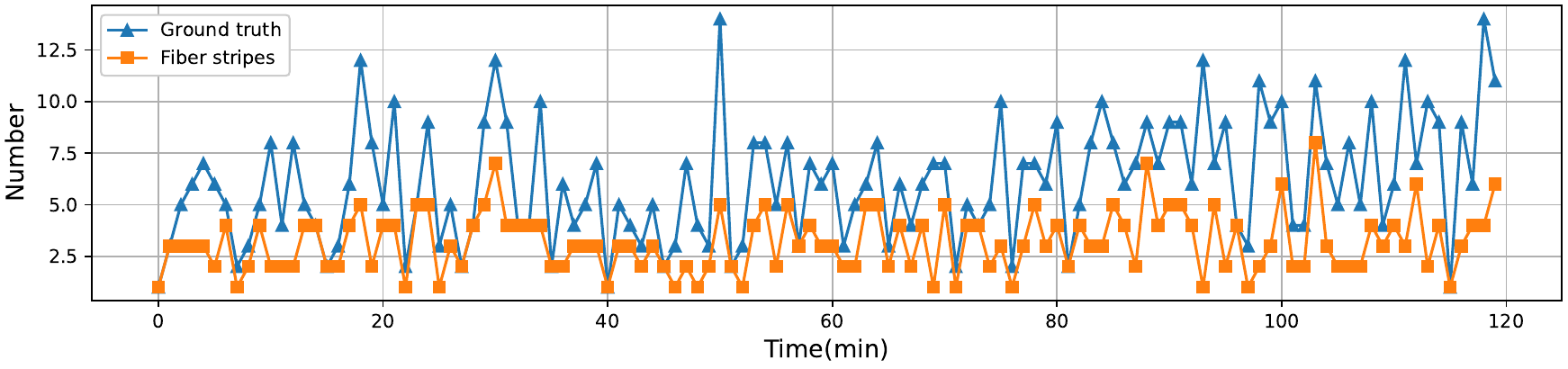}
    }\hfill

\caption{Vehicle statistics counts in Jingjintang expressway dataset including trucks and cars. The orange and blue lines represent fiber stripes and ground truth respectively.}
\label{fica}
\end{figure}

Additionally, we calculated vehicles to assess the capability of DAS systems. Fig. \ref{fica} presents the statistics of vehicle counts for each minute within two hours. Blue lines represent the ground truth counted by the camera, while orange lines represent the fiber stripes counted by the optical fiber. The total number of vehicles within two hours is shown in Table. \ref{highway_fica}. The optical fiber can capture nearly all the information about trucks with an accuracy rate of 94.88\%. Even more accurate to every minute, the fiber stripes can capture almost all large vehicles. However, due to the aforementioned challenges leading to inaccurate optical fiber signals, the counting result for cars is lower than the ground truth only 50\%.

\begin{table}[!t]
    \caption{Comparison of real traffic flow and optical fiber stripe counts within 2 hours of Jingjintang expressway dataset.\label{highway_fica}}
    \renewcommand\arraystretch{1.2}
    \centering
    \begin{tabular}{cccc}
    \hline
    Vehicle type & Fiber stripes & Ground truth & Accuracy rate\\
    \hline
    Truck & 241 & 254 & 94.88\% \\
    Car &  381 & 734 & 51.90\% \\
    Total & 622 & 988 & 62.95\% \\
    \hline
    \end{tabular}
\end{table}

\subsubsection{Jiurui Tunnel Dataset}

\begin{figure}[htbp]
    \centering  
    \includegraphics[width=0.4\textwidth]{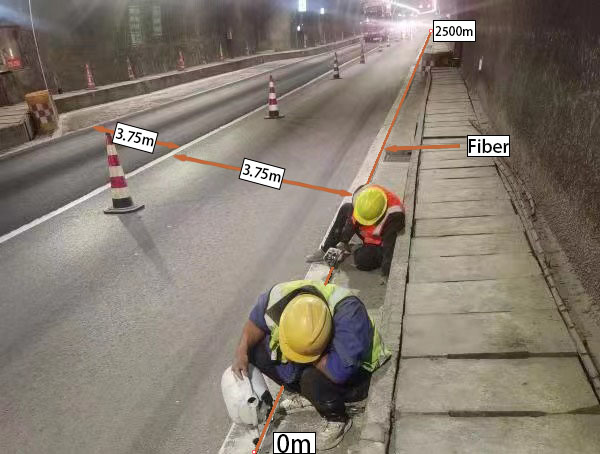}
    \caption{Overhead view of the Jiurui Tunnel, and the optical fiber with a length of 2500m is laid on one side of the road.}
    \label{tunnel_view}
\end{figure}

The total length of the Jiurui tunnel is 2340m with DAS systems through the whole tunnel. The aerial view of detected area is shown in Fig.\ref{tunnel_view}. The optical fiber sampling interval is 0.8m and duration of data collection is 15min. For subsequent analysis, we selected test sections including an 80-meter section and the entire 2000-meter section. The 80-meter section is used for traffic flow analysis and the 2000-meter section contains special cases such as illegal overtaking and congestion, which are used to verify the effectiveness of the proposed algorithms.

\begin{figure}[t!]
    \centering
    \subfigure[2 trucks]{
    \includegraphics[width=1in]{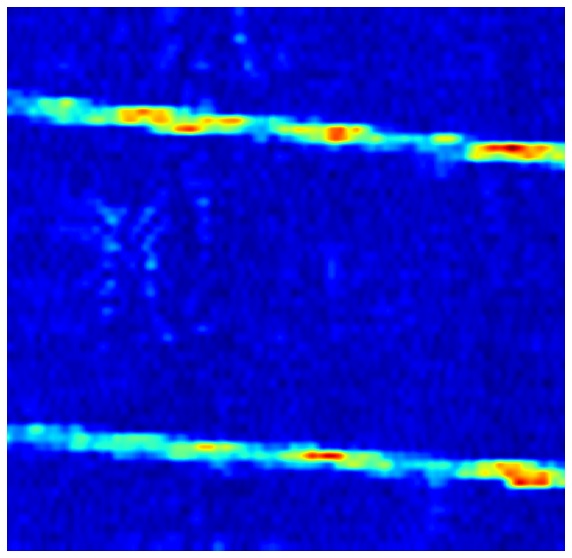}
    }
    \subfigure[4 trucks]{
    \includegraphics[width=1in]{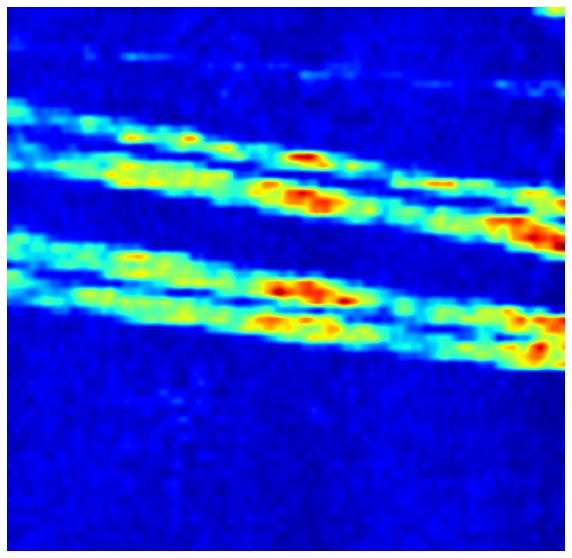}
    }
    \subfigure[2 trucks and 4 cars]{
    \includegraphics[width=1in]{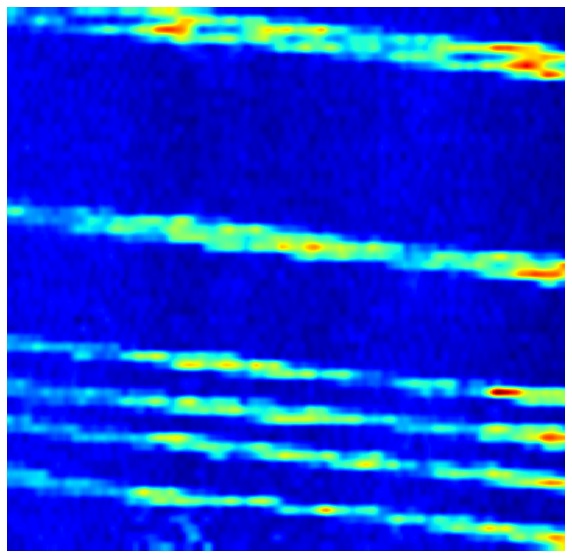}
    }
\caption{Optical fiber waterfall diagrams of vehicles in Jiurui tunnel dataset of 80 meters section. (a) two trucks pass at intervals, (b) four trucks pass at intervals, and (c) two trucks and four cars pass at intervals.}
\label{tunnel_dataset_80}
\end{figure}

Similarly, we draw optical fiber waterfall diagrams to get a more intuitive view of vehicles in Fig. \ref{tunnel_dataset_80}. Compared with Jingjintang expressway dataset, the optical fiber stripes in Jiurui tunnel dataset are more strong and obvious. This distinction arises from the tunnel's configuration, consisting of just two lanes in the same direction, with trucks entering from the outer lane. Consequently, the distinct signals on the waterfall diagram correspond to trucks on the outer lane. Additionally, due to the attenuation effect of the tunnel, the signals from cars in the inner lane are too weak to be adequately recorded. Therefore, the relatively faint signals on the waterfall diagram represent cars in the inner lane.

The Jiurui tunnel dataset of short distance in Fig. \ref{tunnel_dataset_80} comprises 16 sets of optical fiber data, with each set a collection time of 3-4 minutes and a collection distance of 80 meters. Among these, 7 datasets include corresponding video recordings. Comparing these optical fiber stripes in the optical fiber waterfall diagrams with the actual traffic flow in Fig. \ref{tunnel_cafi}, due to signal attenuation caused by the tunnel, the fiber stripes of small vehicles on the outer lane were faint, especially electric vehicles, which could not be adequately captured by DAS systems. This explains the observed variance between optical fiber data acquisition and real traffic flow in tunnel scenario.

\begin{table}[!t]
    \caption{Comparison of real traffic flow and optical fiber stripe counts in Jiurui tunnel dataset of short distance.\label{tunnel_cafi}}
    \centering
    \renewcommand\arraystretch{1.2}
    \begin{tabular}{cccc}
    \hline
    Collection time & Fiber stripes & Ground truth & Accuracy rate\\
    \hline
        14:29-15:32 & 7 & 7 & 100.00\% \\
        15:40-15:43 & 6 & 6 & 100.00\% \\
        15:45-15:48 & 13 & 15 & 86.66\% \\
        16:01-16:04 & 6 & 7 & 85.71\% \\
        16:07-16:10 & 6 & 7 & 85.71\% \\
        16:19-16:22 & 5 & 5 & 100.00\% \\
        16:23-16:26 & 6 & 6 & 100.00\% \\
        \hline
    \end{tabular}
\end{table}

Jiurui tunnel long-distance scenario has low-speed limits and restrictions on lane changes and overtaking. Consequently, heavy traffic flows can lead to traffic congestion. Corresponding waterfall diagrams reveal occurrences of multiple lines intersecting (congestion) and even overlapping (illegal overtaking) due to the specific traffic conditions and regulations enforced in the tunnel.

\subsection{Metrics}
\label{Metrics}

\subsubsection{MSE, PSNR and SSIM}
To quantify the denoising results, we utilize three indices including mean squared error (MSE), the peak signal-to-noise ratio (PSNR) and structural similarity index (SSIM). The formulas are as follows:
\begin{equation}
    {\rm MSE} = \frac{1}{n}\sum\limits_{i = 1}^{n}(y_i - \hat{y}_i)^2,
\end{equation}
\begin{equation}
    {\rm RSNR} = 10{\rm log}\frac{v^2}{\rm MSE},
\end{equation}
where $y_i$ is the original signals, $\hat{y}_i$ is the preprocessed signals, $n$ is total number of sampling points and $v$ is maximum number of image pixels. In general, lower MSE and higher PSNR values imply better reconstruction performance. SSIM is a full-reference evaluation index considering the brightness, contrast and structure of images \cite{wang2004image}. The formulas are as follows:
\begin{equation}
    {\rm SSIM} = [l(y, \hat{y})]^\alpha[c(y, \hat{y})]^\beta[s(y, \hat{y})]^\gamma,
\end{equation}
where $l(y, \hat{y}), c(y, \hat{y}), s(y, \hat{y})$ stand for luminance, contrast and structure respectively. SSIM value ranges from -1 to 1. The closer it is to 1, the better the denoising effect is.

\subsubsection{Traffic indices}

To make statistics on the characteristics of traffic flow, we utilize three fundamental indices: velocity (V), traffic flow (Q), and density (K) on road profile and road segment respectively.

On the road profile, $\rm Q$ is defined as the number of vehicles $\rm N$ passing through a given profile of the road in a given period $\rm T$: $\rm {Q = \frac{N}{T}}$. The average velocity passing through a given point is called time mean speed (TMS). TMS refers to the harmonic average of the location velocity of all vehicles passing through a profile of the road in a certain observation time: $\overline{\rm V}_{\rm {TMS}} = \frac{1}{\frac{1}{\rm N}\sum_{i = 1}^{\rm N}\frac{1}{\rm V_i}}$. Density is expressed as occupancy rate, i.e., time density of vehicles, which represents the ratio of the time occupied by vehicles passing a certain road profile to the observation time in a certain observation time: $\rm K = \frac{\rm Q}{\overline{V}_{\rm {TMS}}}$.

On the road segment, $\rm K$ is the spatial density set of vehicles, that is, the number of vehicles $\rm N$ existing per unit road length $\rm L$ at a certain moment is $\rm K = \frac{\rm N}{\rm L}$. The average velocity of the vehicles in a given section is known as space mean speed  (SMS). SMS refers to the arithmetic average of all vehicle instantaneous velocities in a certain observation section: $\overline{\rm V}_{\rm {SMS}} = \frac{\sum_{i = 1}^{\rm N}\rm V_i}{\rm N}$. Traffic flow refers to the ratio of the total distance of all vehicles to the accumulation of space and time in a certain period, $\rm Q = \rm K\overline{\rm V}_{\rm {SMS}}$.

\subsection{Analysis}
\label{analysis}

\begin{figure}[!t]
    \centering
    \subfigure[Without denoising.]
    {\includegraphics[width=2in]{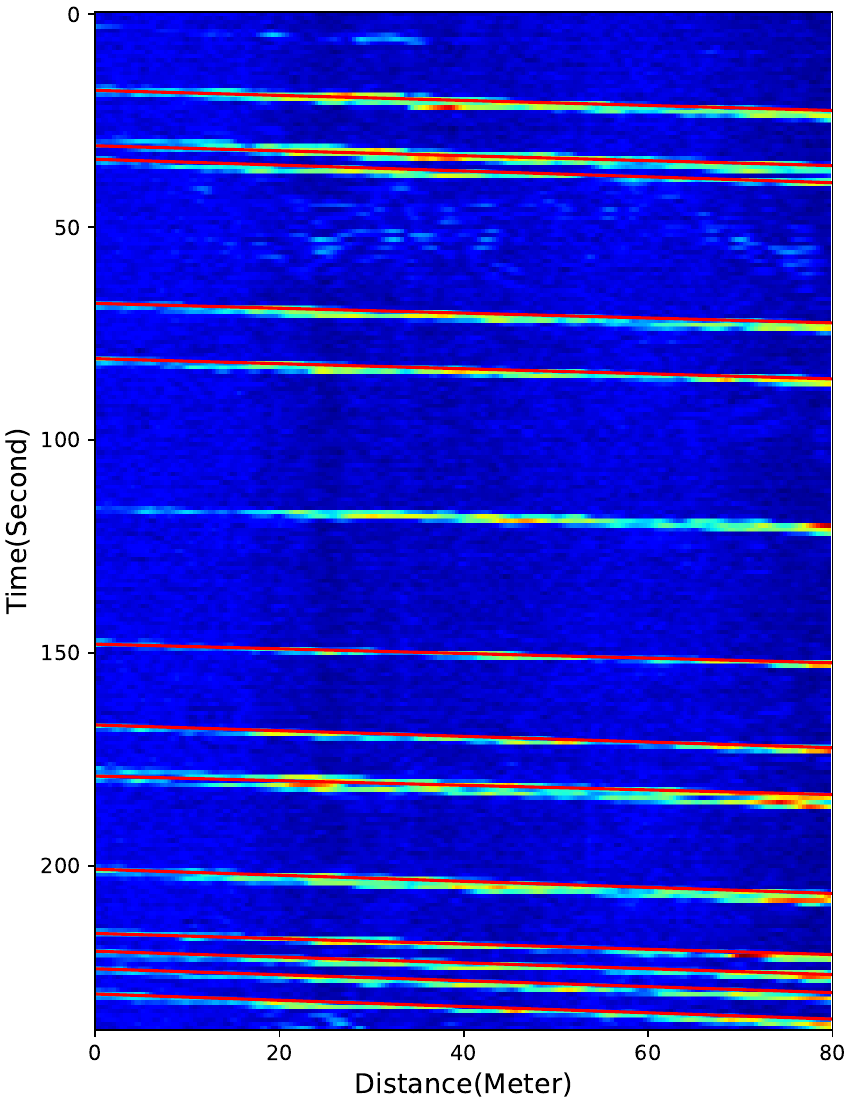}}
    \subfigure[Soft threshold denoising.]
    {\includegraphics[width=2in]{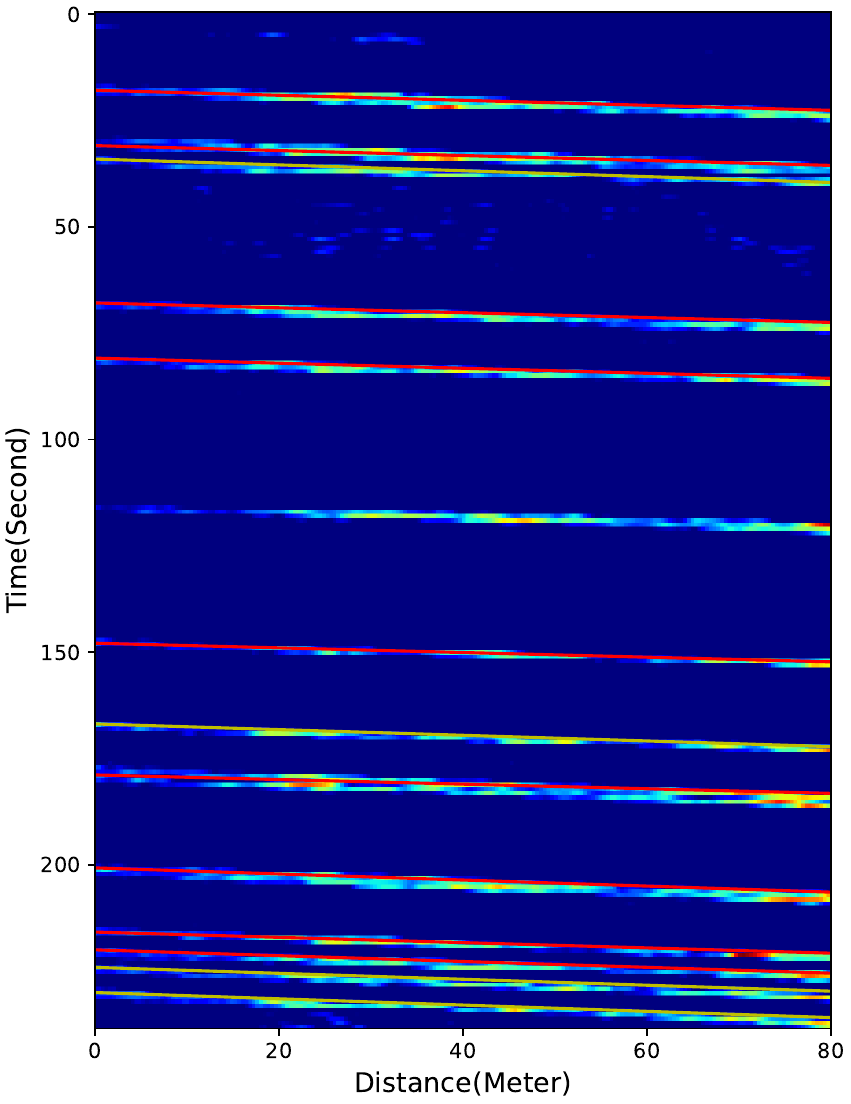}}
    \subfigure[Improved threshold denoising.]
    {\includegraphics[width=2in]{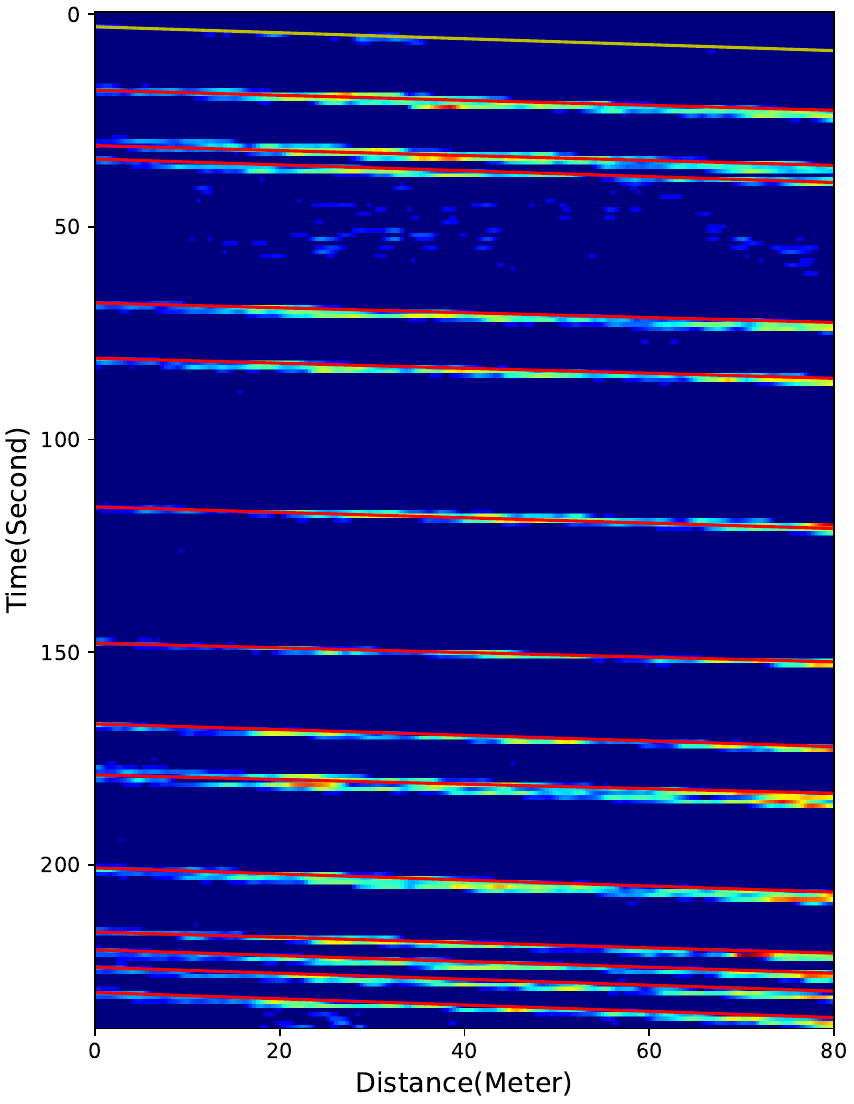}}
    \caption{An example of vehicle tracking results in Jiurui tunnel dataset of short distance. The red lines represent the extracted trajectory of trucks while the yellow line represents a car. (a) Trajectory without denoising, (b) trajectory using soft threshold denoising, and (c) trajectory using improved threshold denoising.}
    \label{trajectory_tunnel_short}
\end{figure}

We tested the proposed algorithms on real tunnels and cities. In tunnels, the wavelet denoising threshold $\lambda = 0.15$ and $a = 0.5$ in Eq. (\ref{wavelet}). Butterworth low-pass filter was applied to the first col signals with order $N = 1$ and normalized cut-off frequency $W_n = 0.5$. Then we extracted signal peaks using the minimum height of 0.06. In the phase of vehicle tracking, the initial velocity interval is $v_{\min}^{(1)} = 60$ km/h to $v_{\max}^{(1)} =120$ km/h. The vehicles with an average peak value above 0.5 and width above 3 were trucks and the rest were cars. Fig. \ref{trajectory_tunnel_short} depicts ablation experiments regarding different denoising algorithms. In these vehicle trajectory waterfall diagrams, each vehicle is uniquely identified with an individual ID. We conducted quantitative comparative experiments on the effectiveness of the denoising algorithm. As shown in Table. \ref{index}, improved threshold denoising method has smaller MSE, larger PSNR and SSIM, indicating better results.

\begin{table}
	\centering
    \renewcommand\arraystretch{1.2}
	\caption{Quantitative indexes of different denoising algorithms.}
    \label{index}
	\begin{tabular}{cccc}
		\hline  
         Method & MSE & PSNR & SSIM\\
        \hline
            Soft wavelet & 0.0568 & 73.0321 & 0.8645\\
            Improved wavelet & \textbf{0.0356} & \textbf{77.0969} & \textbf{0.9161}\\

        \hline
	\end{tabular}
\end{table}

Table. \ref{tunnel_fial} present results of algorithm test counts and fiber stripe counts in Jiurui tunnel dataset. Among them, there were 143 instances of fiber stripes, and the algorithm's detection results yielded an accuracy of 90.90\%. Even on certain road sections, the accuracy rate can reach 100\%. 

\begin{table}[!t]
    \caption{Comparison of algorithm test counts and optical fiber stripe counts in 7 sets of Jiurui tunnel dataset.\label{tunnel_fial}}
    \renewcommand\arraystretch{1.2}   
    \centering
    \begin{tabular}{cccc}
    \hline
    Collection time & \makecell[l]{Algorithm test \\counts}& \makecell[l]{Fiber stripes\\ counts}  & Accuracy rate\\
        \hline
        14:29-15:32 & 5 & 7 & 71.42\% \\
        15:40-15:43 & 6 & 6 & 100.00\% \\
        15:45-15:48 & 11 & 13 & 84.61\% \\
        16:01-16:04 & 5 & 6 & 83.33\% \\
        16:07-16:10 & 5 & 6 & 83.33\% \\
        16:19-16:22 & 5 & 5 & 100.00\% \\
        16:23-16:26 & 6 & 6 & 100.00\% \\
        \hline
    \end{tabular}
\end{table}

\begin{figure}[t!]
    \centering
    \quad
    \subfigure[Hough transform.]{
    \includegraphics[width=2in]{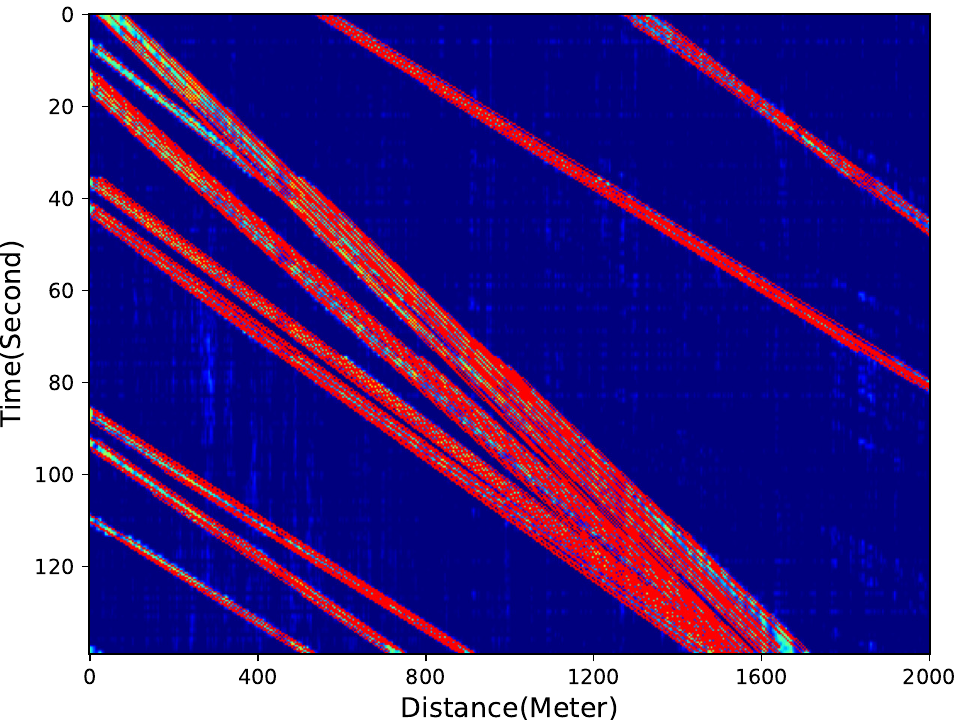}
    \includegraphics[width=2in]{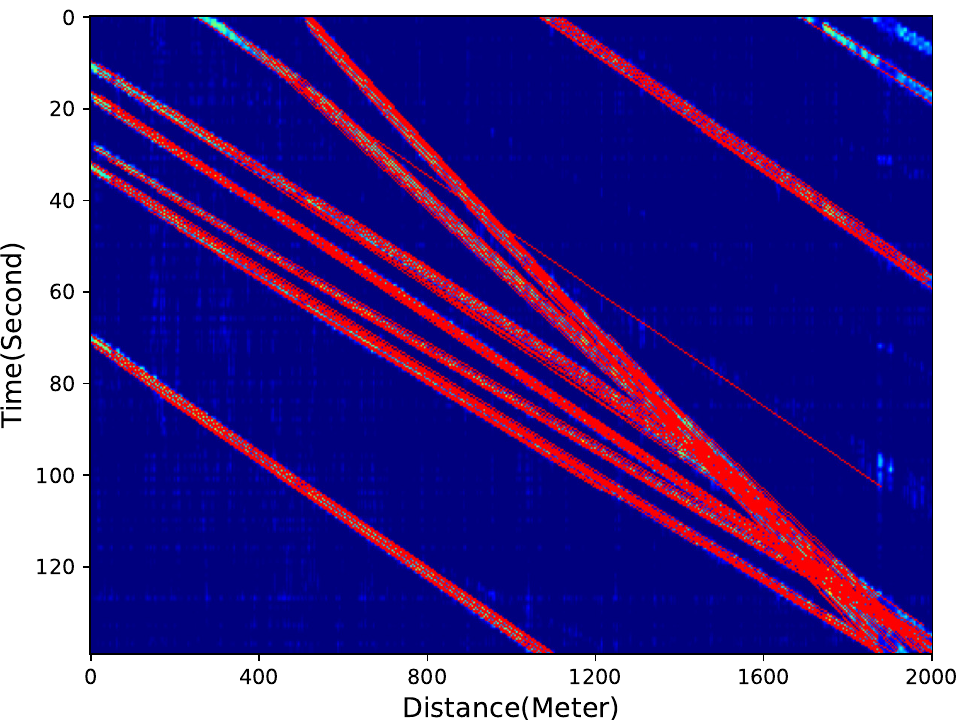}
    \includegraphics[width=2in]{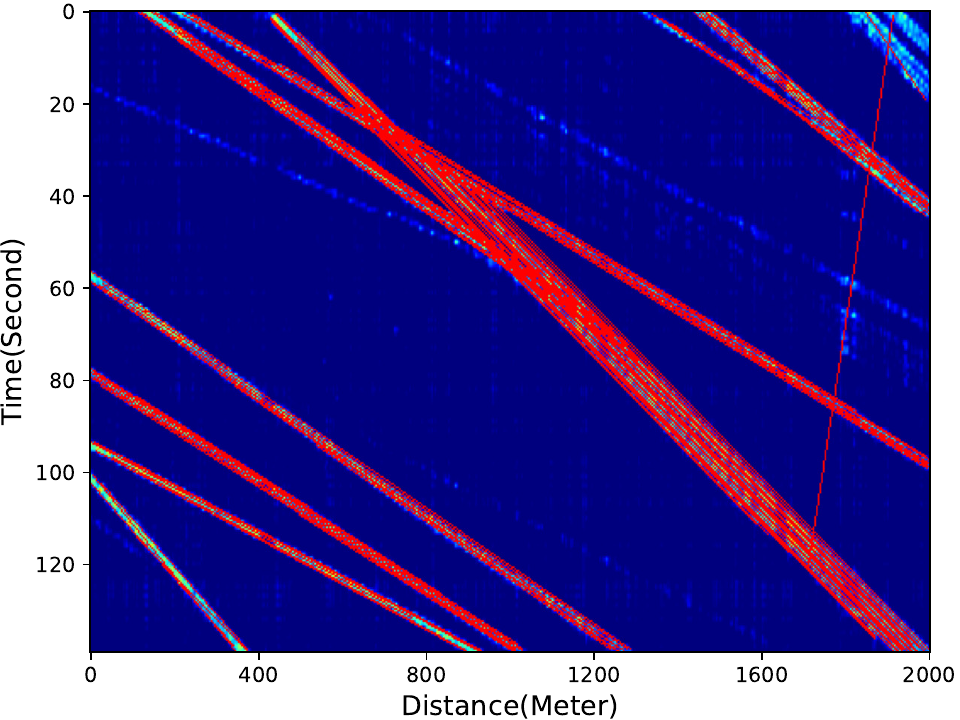}
    }

    \quad
    \subfigure[Radon transform.]{
    \includegraphics[width=2in]{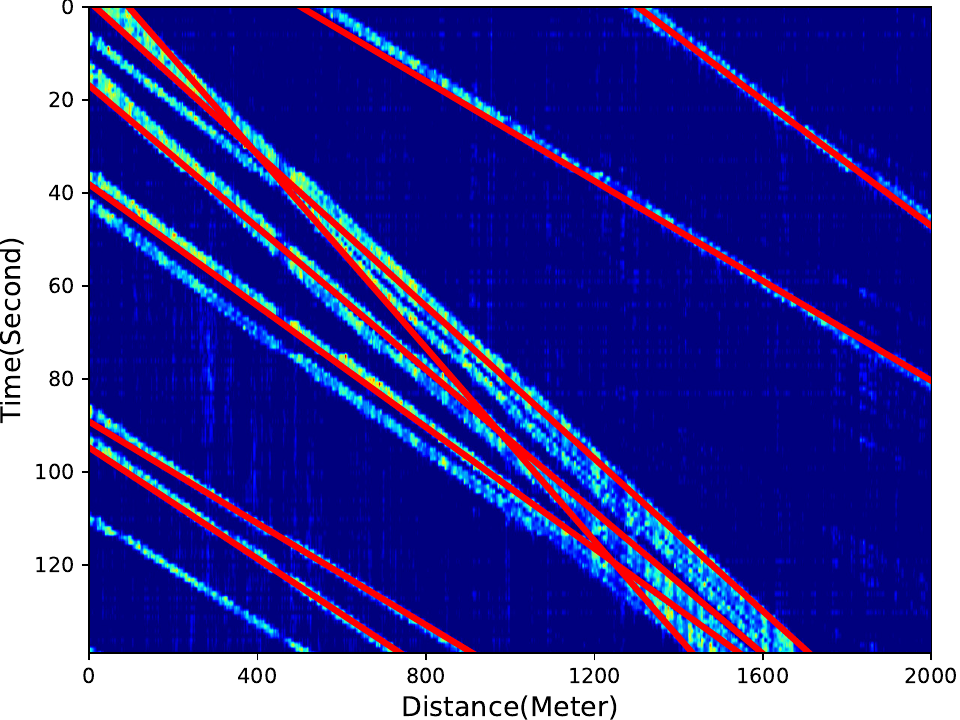}
    \includegraphics[width=2in]{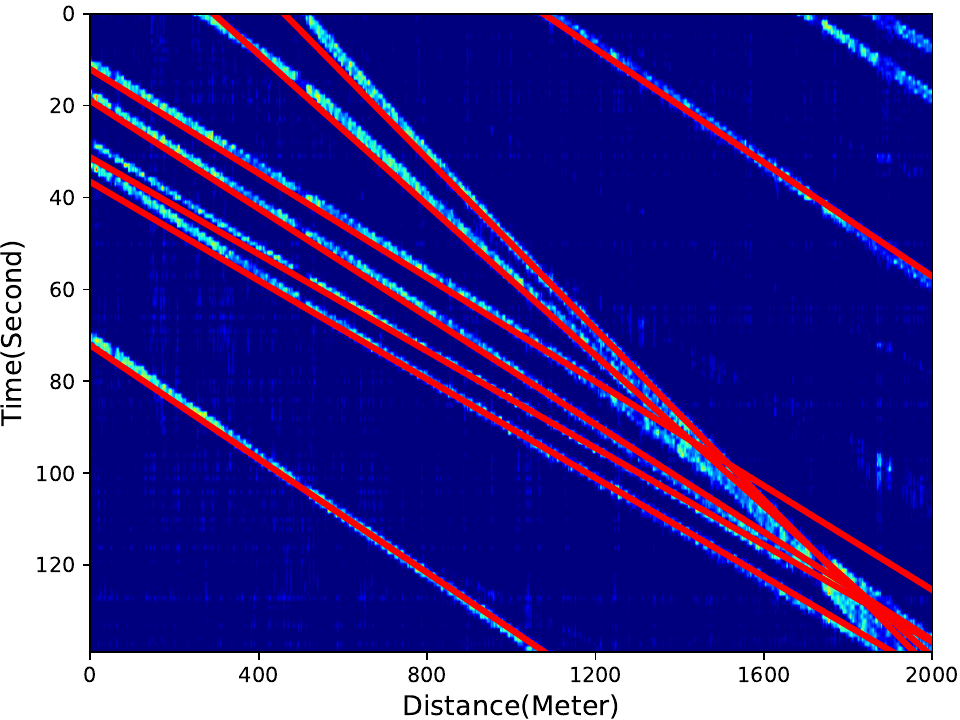}
    \includegraphics[width=2in]{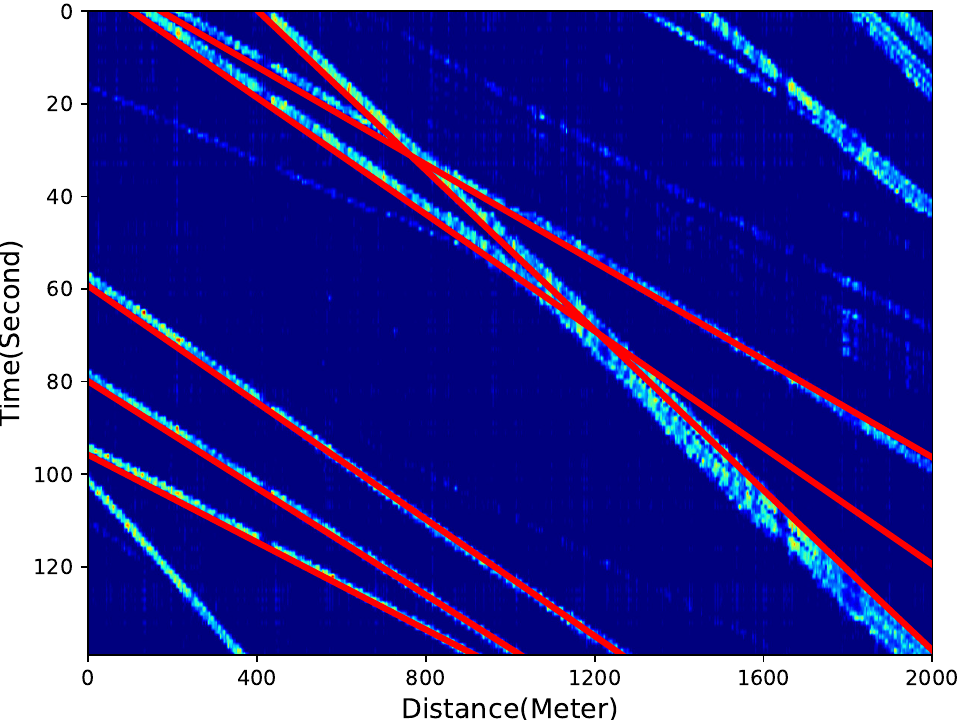}
    }

    \quad
    \subfigure[Our proposed method.]{
    \includegraphics[width=2in]{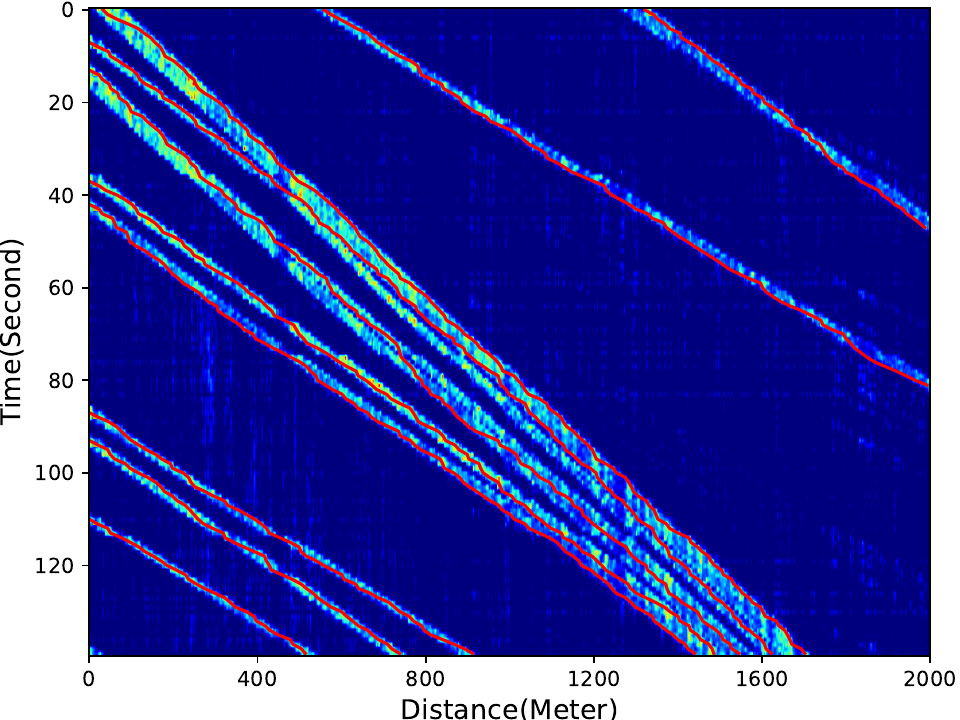}
    \includegraphics[width=2in]{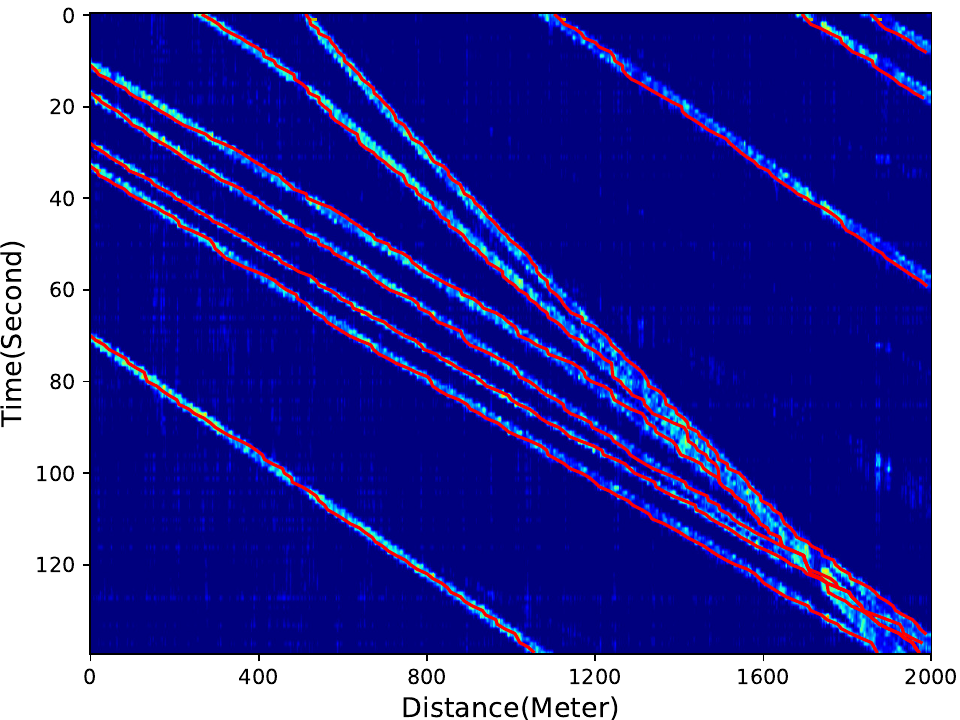}
    \includegraphics[width=2in]{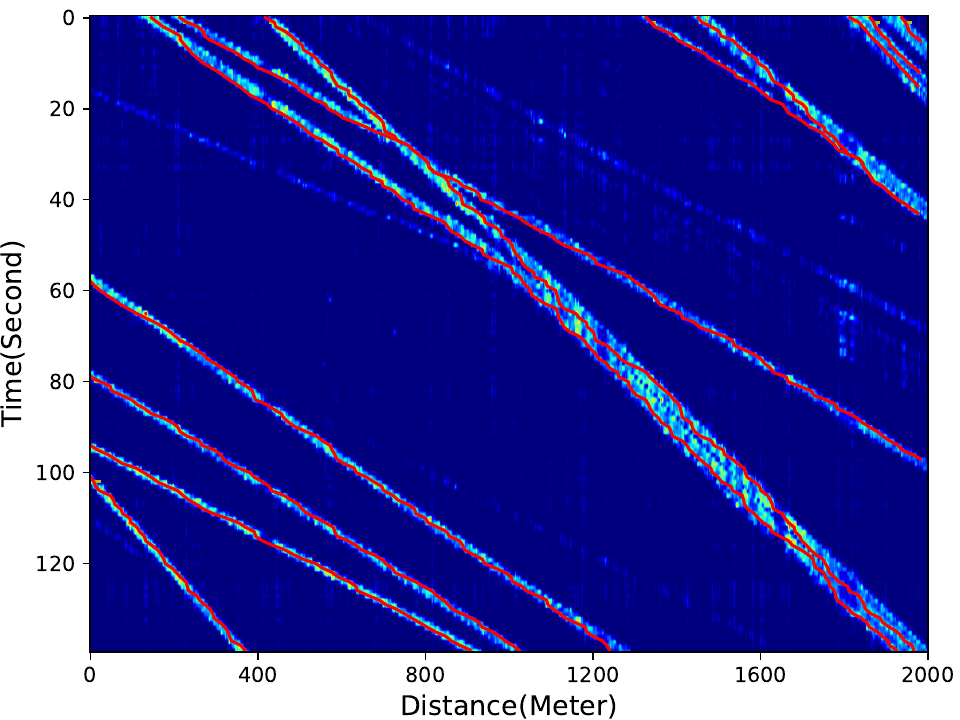}
    }

\caption{Vehicle trajectory extraction results. (a) Hough transform, (b) Radon transform, and (c) our proposed method.}
\label{trajectory_tunnel_long}
\end{figure}

Furthermore, after employing the same preprocessing steps, we carried out comparative experiments with Hough transform \cite{wiesmeyr2021distributed, catalano2021automatic} and Radon transform \cite{wang2021vehicle} under complicated events like congestion and overtaking events. As depicted in Fig. \ref{trajectory_tunnel_long}, Hough transform identifies a single vehicle trajectory line as multiple lines; Radon transform overlooks detecting lines with short driving distances. Our algorithms can effectively distinguish and separate individual vehicles, showing superior consistency and accuracy even when lines intersect or overlap.

\begin{figure}[t!]
    \centering
    \includegraphics[width=6in]{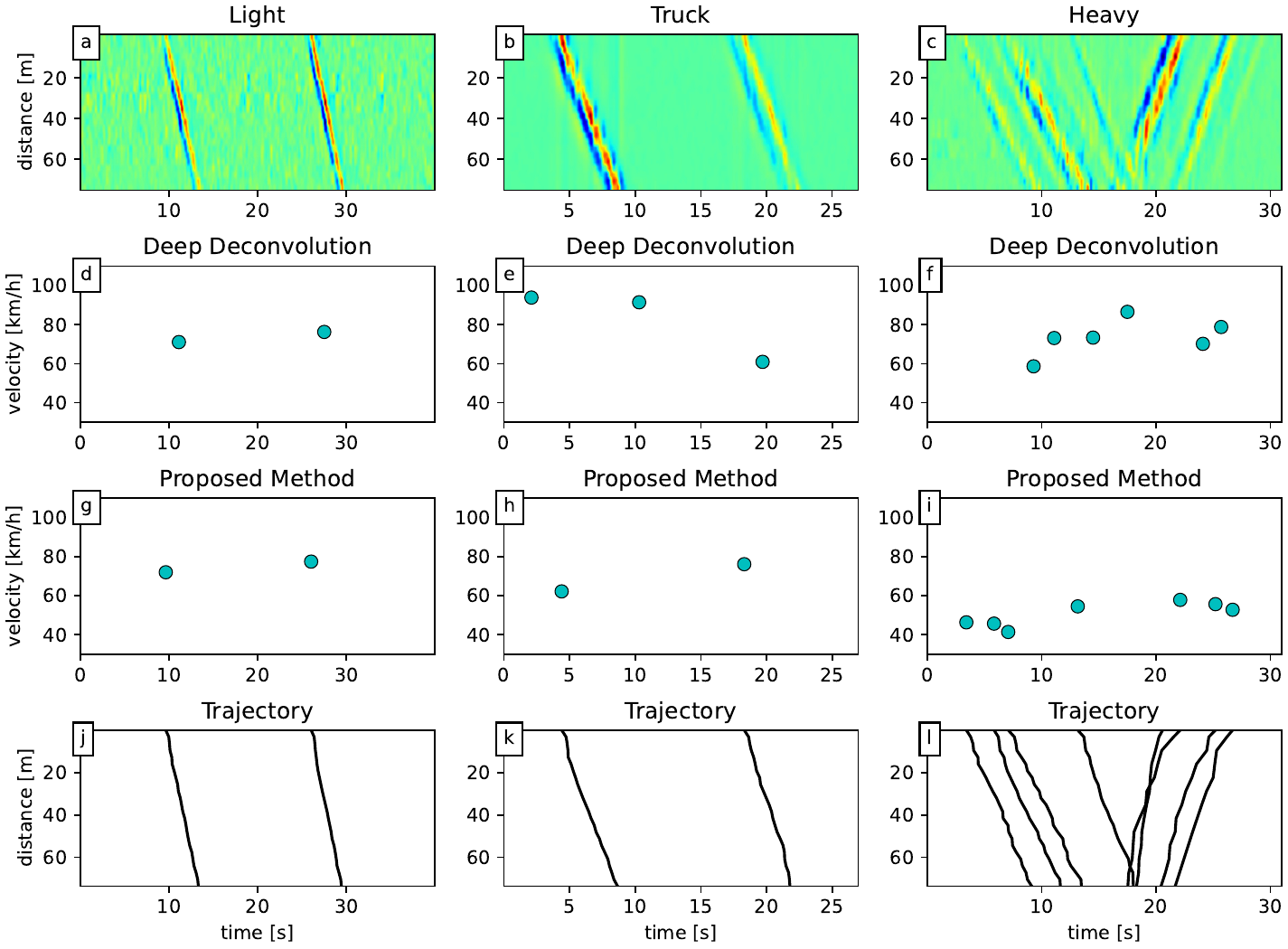}
\caption{Three examples of different traffic conditions, ordered by increasing difficulty. The top panels (a, b, c) show the original DAS recordings, the panels (d, e, f) and (g, h, i) show the results of deep deconvolution and proposed algorithms respectively. Furthermore, our algorithm can track the whole trajectories, as shown in panels (j, k, i).}
\label{deep}
\end{figure}

Considering that our experiments are only carried out on highways, we conduct comparative experiments on public datasets of normal urban traffic in \cite{van2022deep}. Besides, we compare our proposed algorithm with deep deconvolution and MUSIC beamforming algorithms\cite{van2022deep} to further prove the effectiveness of our proposed algorithms. As shown in Fig. \ref{deep}, we selected three different representative fiber signals in datasets of \cite{van2022deep} for experiments. It is worth mentioning that our algorithm can better extract the entry time and velocity information of vehicles, and can track different vehicles throughout the whole process.

In addition to vehicle trajectory extraction, our approach can perform comprehensive statistical analyses, including velocity, flow and density on both road profiles and road segments. For example, based on the extraction results of structured vehicle data from the dataset in Fig. \ref{trajectory_tunnel_short} of the tunnel scenario, we can specify a certain road profile or a certain segment and make traffic flow information statistics like Table. \ref{tunnel_fsd}.

\begin{table}
	\centering
    \renewcommand\arraystretch{1.2}
	\caption{Some traffic flow information statistics.}
    \label{tunnel_fsd}
	\begin{tabular}{ccccccc}
		\hline  
        & \multirow{2}*{Time} & \multirow{2}*{Position} & \multicolumn{4}{c}{traffic flow information statistics}\\
            \cline{4-7}
            & & & \makecell[l]{Vehicle \\counts} & \makecell[l]{Average \\velocity} & Flow &Density\\
            \hline
            \multirow{4}*{\makecell[l]{Road \\profile}} & 0-1min & 40m & 4 & 56.38km/h & 0.06 & 0.0011\\
                                        & 1-2min & 40m & 3 & 59.93km/h & 0.05 & 0.0008\\
                                        & 2-3min & 40m & 2 & 59.27km/h & 0.03 & 0.0005\\
                                        & 3-4min & 40m & 6 & 54.07km/h & 0.10 & 0.0018\\
            \cline{1-7}
            \multirow{4}*{\makecell[l]{Road \\segment}} & 0-40m & 30s & 1 & 55.97km/h & 0.55 & 0.0100\\
                                        & 0-80m & 30s & 1 & 55.97km/h & 0.27 & 0.0050\\
                                        & 0-40m & 150s & 1 & 65.76km/h & 0.65 & 0.0100\\
                                        & 0-80m & 150s & 1 & 65.76km/h & 0.32 & 0.0050\\
        \hline

	\end{tabular}
\end{table}

\section{Perspectives}
\label{perspectives}

Despite the advantages of DAS systems, such as low cost, precise position, and corrosion resistance, it does have inherent limitations. As explained in Section \ref{Jingjintang Expressway Dataset}, there are several complex traffic situations difficult to deal with, including traffic flow with cars and trucks, multi-vehicle parallel, and opposite interference. Having described the potential of DAS systems to promote ITS, we further explore the possibility of DAS systems and camera systems integration to achieve better results.

The video-based multi-object tracking (MOT) algorithm is divided into two stages: object detection and object tracking. In the object detection stage, we utilized deep learning-based YOLOv8 \cite{reis2023real} algorithm and fine-tuned it on the UA-DETRAC dataset \cite{wen2020ua}. In the object tracking stage, we applied the BoT-SORT algorithm \cite{aharon2022bot} for subsequent object tracking, which is a multi-object tracking algorithm that uses a two-stage Kalman filter and object feature matching. By integrating the YOLOv8 and BoT-SORT algorithms, we can obtain vehicle tracking results from video-based systems. Further, we transform the video perspective projection to the aerial perspective to compare with fiber-based systems using perspective transformation \cite{haralick1980using}. The perspective matrix here is:
\begin{equation}
\begin{pmatrix}
         -0.051 & -0.007  & 75.813 \\
        0 & -0.041  & 44.298 \\
        0 & 0.007 & 1
         \end{pmatrix}
\end{equation}

\begin{figure}[!t]
    \centering
    \includegraphics[width=6in]{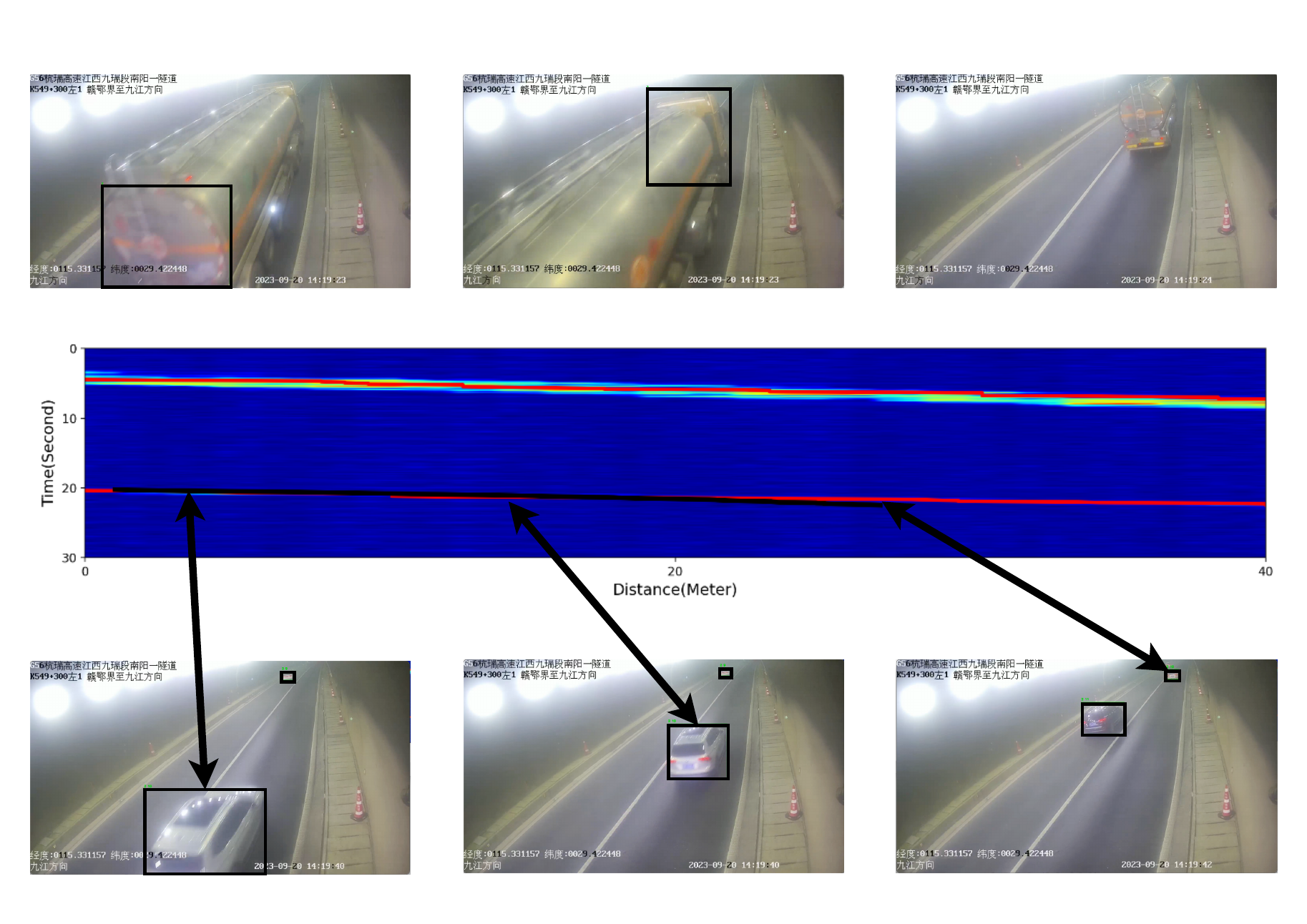}
    \caption{Fiber-based and vision-based tracking results. In the waterfall diagram in the middle, the two red lines represent the DAS tracking results of the two vehicles. A black line represents the video tracking track of the following line of vehicles after video detection and tracking, and describes the corresponding relationship between the positions of vehicles in the three pictures.}
    \label{camera_fiber}
\end{figure}

As shown in Fig. \ref{camera_fiber}, two vehicles passed through within 30 seconds with a tanker in front and a car behind. The two red lines represent the fiber-based tracking results of the two vehicles. Due to the length of the tanker, only a few frames can be detected, making it impossible to track the entire trajectory in vision-based system. Only the car can be tracked throughout the process. After perspective transformation, the car's trajectory is projected onto the optical fiber diagram, as shown by the black line, with bidirectional arrows indicating the corresponding spatial relationship between the video and the fiber images. It is worth mentioning that due to the continuous laying of optical fiber, vehicles can be continuously tracked, unrestricted by the fixed monitoring angle of cameras. Since DAS systems can only provide accurate vehicle location information rather vehicle and road conditions visualization, we can consider providing more accurate information through the video perspective, such as lane, model, license plate, color and other information. Therefore, in future work, the combination of video-based and fiber-based systems can be furthur explored to promote the further development of ITS.

\section{Conclusion}
\label{Conclusion}
    
In this paper, we have demonstrated the feasibility of employing DAS systems for traffic monitoring. Considering the challenges when facing vehicle complex behaviors like congestion and illegal overtaking, we designed a series of signal processing and feature extraction to achieve multi-vehicle tracking. The data preprocessing utilizes a wavelet threshold denoising algorithm combining hard and soft threshold functions. Based on preprocessed optical fiber waterfall diagrams, we propose a vehicle position detection algorithm using the Butterworth low-pass filter and peaks location search. Standard indices MSE, PSNR and SSIM were employed for quantitative analysis to enhance data comprehension. Then a line-by-line match algorithm is proposed for vehicle trajectory extraction to achieve overall multi-vehicle tracking. Using the extracted vehicle information, the velocity, traffic flow and density of any specified road profile and road segment can be further calculated. Experiments were carried out in highway, tunnel and city scenarios. In the simple section of tunnel scenario, the algorithm's detection results yielded an accuracy of 90.90\%. In some complex sections, compared with the classical Hough and Radon transform-based methods and MUSIC beamforming algorithm, the visualized results of our proposed methods demonstrate advantages in tracking vehicles and measuring speed in long-distance.


However, there are certain deficiencies and areas for improvement in this research, including:

(1) \textbf{Handing heavy traffic and parallel passages:} Our proposed vehicle detection method works effectively for simple traffic scenarios. However, in cases of heavy traffic with large and small vehicles passing in parallel, improving how we distinguish vehicle counts and precisely identify vibration signal positions is necessary.

(2) \textbf{Dependency on vehicle starting points:} The vehicle detection algorithm in this paper heavily relies on the starting points of vehicles. This approach may lead to missed detections when only using optical fiber data. In the era of multi-sensor fusion, integrating cameras could help determine the precise starting positions of vehicles, thereby improving trajectory accuracy.

(3) \textbf{Considering complex traffic scenarios:} To deploy fiber-assisted intelligent transportation successfully, we also need to address more complex vehicle behaviors in real highway conditions, such as overtaking and stopping, to provide early warnings of emergencies.

Overall, our study demonstrates the great potential of DAS applications in traffic monitoring. In future work, we expect to integrate the advantages of multiple sensors (such as fibers, videos and cameras) and further serve intelligent transport systems.

\section*{Declaration of Competing Interest}

The authors declare that they have no known competing financial interests or personal relationships that could have appeared to influence the work reported in this paper.

\printcredits

\section*{Acknowledgments}

This research was funded by Tianjin Yunhong Technology Development (Grant number: 2021020531).

\bibliographystyle{elsarticle-num-names.bst}


\bibliography{reference}

\end{document}